\documentclass[preprint2]{aastex}

\begin{document}

\title{Evolution of Star Forming and Active Galaxies in Nearby Clusters}
\author{Neal A. Miller\altaffilmark{1,2}} 
\affil{NASA Goddard Space Flight Center \\ UV/Optical Branch, Code 681\\Greenbelt, MD \ 20771}
\email{nmiller@stis.gsfc.nasa.gov}

\and

\author{Frazer N. Owen}
\affil{National Radio Astronomy Observatory\altaffilmark{3}, P.O. Box O, \\ Socorro, New Mexico 87801}
\email{fowen@aoc.nrao.edu}

\altaffiltext{1}{Visiting Astronomer, Kitt Peak National Observatory, National Optical Astronomy Observatories, which is operated by the Association of Universities for Research in Astronomy, Inc. (AURA) under cooperative agreement with the National Science Foundation.}

\altaffiltext{2}{Based in part on observations obtained with the Apache Point Observatory 3.5-meter telescope, which is owned and operated by the Astrophysical Reseach Consortium.}

\altaffiltext{3}{The National Radio Astronomy Observatory is a facility of the National Science Foundation operated under cooperative agreement by Associated Universities, Inc.}

\begin{abstract} 
We have used optical spectroscopy to investigate the active galaxy populations in a sample of 20 nearby Abell clusters. The targets were identified on the basis of 1.4 GHz radio emission, which identifies them as either AGN or galaxies forming stars at rates comparable to or greater than that of the Milky Way. The spectra were used to characterize the galaxies via their emission and absorption features. 

The spectroscopy results reveal a significant population of star forming galaxies with large amounts of nuclear dust extinction. This extinction eliminates bluer emission lines such as [OII] from the spectra of these galaxies, meaning their star formation could easily be overlooked in studies which focus on such features. Around $20\%$ of the cluster star forming galaxies have spectra of this type. The radial distributions of active galaxies in clusters show a strong segregation between star forming galaxies and AGN, with star forming galaxies broadly distributed and AGN preferentially in the cluster cores.  The radial distribution of the dusty star forming galaxies is more centrally-concentrated than the star forming galaxies in general, which argues that they are a consequence of some cluster environmental effect. Furthermore, we note that such galaxies may be identified using their $4000\mbox{\AA}$ break strengths.

We find that discrepancies in reported radio luminosity functions for AGN are likely the result of classification differences. There exists a large population of cluster galaxies whose radio fluxes, far-infrared fluxes, and optical magnitudes suggest their radio emission may be powered by stars, yet their spectra lack emission lines. Understanding the nature of these galaxies is critical to assessing the importance of AGN in the radio luminosity function at low luminosities. We also find that regardless of this population, the crossover point where the radio luminosity function is comprised equally of star forming galaxies and AGN occurs at lower luminosities in clusters than the field. This is likely a simple consequence of the reduction in star formation in cluster galaxies and the morphological mix in clusters compared to the field.

\end{abstract}

\section{Introduction}

The role of environment in the evolution of galaxies has been a central question in extragalactic astronomy. One of the stronger pieces of observational evidence for galaxy evolution in the environment of rich clusters of galaxies is the Butcher-Oemler effect \citep{butc1978,butc1984}. Subsequent photometric work to examine this effect and its causes has confirmed its existence with larger samples \citep{rako1995,lubi1996,marg2000,marg2001} and revealed more evidence for evolution in cluster populations. \citet{dres1997} used HST imaging of galaxies in distant clusters to confirm that the morphology-density relationship still holds, but with important differences. Specifically, the fraction of S0 galaxies is several times lower in distant clusters, arguing that while elliptical galaxies formed in the distant past the S0 galaxies seem to have formed after cluster virialization.

Spectroscopic studies motivated by the Butcher-Oemler effect have also produced evidence for evolution in clusters \citep[e.g.,][]{dres1983,couc1987,dres1992,dres1999}. \citet{dres1983} uncovered an unusual population of galaxies in distant clusters, dubbed ``E+A'' galaxies, and noted that their frequency in clusters increased at higher redshift. While these galaxies are more common in clusters than the field at high redshift, their presence in the field (even locally) implies that they can not be explained solely by environmental effects exclusive to cluster environments \citep{zabl1996}. The spectroscopic catalog of \citet{dres1999} noted the presence of an additional important population of galaxies, those which exhibited strong Balmer absorption with slight emission of the star formation indicator, [OII] $\lambda3727$. In general, the frequency of these galaxies mimics that of the E+A galaxies, being more common at higher redshift. \citet{pogg1999} applied evolutionary models to investigate these galaxies and argued that they were most likely dusty starbursts, and drew parallels to galaxies in the nearby universe \citep{liu1995,pogg2000}. Further models confirm that these dusty starbursts can be the precursors to the post-starburst galaxies \citep{shio2001}.

Ironically, the exciting evidence for evolution in clusters has diverted attention from nearby clusters. Since clusters in the nearby universe seem to exhibit less activity (in the sense of harboring galaxies with strong current or recent star formation), they are often overlooked in favor of their predecessors at intermediate redshift \citep[although see][]{cald1993,cald1997,rose2001}. This is unfortunate, since nearby clusters can more easily be investigated in greater detail. Not only can such investigations provide accurate benchmarks for studies of clusters at higher redshift, but they can influence future high redshift studies by uncovering trends not easily apparent given the practical limits of present-day telescopes and detectors.

In this paper, a large sample of active galaxies drawn from nearby clusters is used to assess galaxy evolution in the cluster environment. The galaxies are radio selected, with 1.4 GHz luminosities consistent with that of the Milky Way or greater. Radio emission is an excellent indicator of activity, in the form of either star formation or active galactic nuclei \citep{cond1992}. The sample is primarily that identified in \citet{mill2001}, augmented by comparable radio observations of two additional clusters. Cluster membership of the radio galaxies has been achieved via optical spectroscopy, producing a net sample of 411 confirmed members in 20 Abell clusters (plus an additional 12 potential members for which velocities were not obtained). Much of the data used in this paper correspond to quality long-slit spectra of a large subset of the confirmed cluster members. Classification of the galaxies is performed via this spectroscopy, including line ratio diagnostics to characterize the activity (star formation, AGN, or both) and assess the importance of dust extinction. The distributions of these various classes of galaxies are then compared to elucidate the role of environment on cluster galaxy evolution.

Specifically, this paper examines the sample collectively to explore several issues of galaxy evolution in the cluster environment. First, what types of active galaxies are seen in nearby clusters, and in what numbers? In part, this amounts to understanding the relative importance of star formation and AGN in nearby clusters. Within the star forming galaxies, the spectra also provide a means to assess the importance of dust extinction. Second, what can the distributions of these active galaxies tell us about evolutionary mechanisms in clusters? Because the presence of radio emission implies recent activity, the locations of the active galaxies are tied to their local environment. Consequently, the distributions will elucidate the importance of various evolutionary models. Lastly, what can the sample tell us about the composition of the radio luminosity function and its application to studies of star formation at higher redshift? While these issues are addressed using the sample collectively, in future papers we will make cluster-by-cluster comparisons in order to shed light on issues of specific cluster parameters and their effects on member galaxies.

The paper is organized as follows: In Section \ref{sec:data}, the data are summarized. This includes an overview of the sample and observations, as well as the procedures used to classify the cluster galaxies on the basis of their optical spectra. The different classes of radio galaxies are then investigated in Section \ref{sec:analysis}. This includes an assessment of the importance of dust extinction and an investigation of how the different classes of radio galaxies are distributed in the clusters. Section \ref{sec:rlf} presents the cluster radio luminosity function, broken down by activity class. This breakdown also leads to a deeper exploration of the assigned activity classes, in particular the nature of those galaxies with absorption-line spectra.  The implications of all these results are discussed in Section \ref{sec:discuss}, followed by a brief summary of the conclusions in Section \ref{sec:conclude}. Unless otherwise noted, we have assumed $H_o = 75$ km s$^{-1}$ Mpc$^{-1}$ and $q_o = 0.1$ throughout.

\section{Data}\label{sec:data}

\subsection{The Sample}

The sample is primarily that of \citet[][hereafter Paper I]{mill2001}. This paper used the National Radio Astronomy Observatory VLA Sky Survey \citep[NVSS;][]{cond1998} to identify active galaxies in 18 nearby Abell clusters \citep{abel1989}. The radio galaxies were identified down to an optical magnitude of $M_R=-20$ \citep[$m^*+2$ from][]{owen1989} over an area extending from the cluster centers out to projected radii of 3 Mpc. This is 1.5 times the classical Abell radius, meaning the identified galaxies will better trace the transition from the rich cluster environment to the field. Examination of published velocities and a comprehensive program of optical spectroscopy produced lists of cluster members and non-cluster galaxies seen in projection.

The value of such a radio-selected sample is that it is an unbiased method to identify galaxies, be they star forming, AGN, or some combination thereof. At higher luminosities, the 1.4 GHz radio luminosity function (RLF) is dominated by AGN \citep{mach2000,sadl2002}. At lower luminosities, the composition of the RLF shifts to mainly star forming galaxies with the luminosity being directly related to the star formation rate (SFR). This is especially powerful since the radio emission is unaffected by dust extinction. In the case of the Paper I sample, the NVSS flux limit corresponds to SFRs comparable to that of the Milky Way for the identified cluster galaxies. In total, the identified radio galaxies represent $\sim15\%$ of all optical galaxies in the prescribed magnitude range. Due to the fixed flux limit of the NVSS and the range in cluster redshifts, applying a uniform radio luminosity cutoff to all clusters reduces this figure to $\sim9\%$. As would be expected, the radio detection rate is dependent on the optical magnitude of the galaxies, reaching $\sim30\%$ for galaxies brighter than $M_R=-22$ and dropping to a few percent for galaxies fainter than $M_R=-21$.

In addition to providing velocities for assessment of cluster membership, optical spectra are useful in characterizing the galaxies. Analysis of the emission and absorption features of the galaxies is used to evaluate whether the galaxies possess active nuclei or are dominated by star formation. Furthermore, in the case of star forming galaxies the spectra may be used to assess the amount of dust extinction present.

The clusters comprising the sample from Paper I were selected based on their proximity ($cz \leq 10,000$ km s$^{-1}$). Two additional clusters have been added for this study on the basis of their being prime candidates for nearby cluster-cluster mergers. \citet{dwar1999} and \citet{owen1999} noted that such large-scale mergers might have profound effects on constituent galaxies, and hence it was desirable to add examples of such systems to the clusters being studied. Consequently, Abell 2255 and Abell 2256 are included in this paper. Lying at slightly higher redshifts than the rest of the cluster sample ($z\sim0.08$ and $z\sim0.06$, respectively), the flux limit of the NVSS was too high to explore star forming galaxies in the clusters. As a result, we have observed these clusters using mosaics of pointings with the VLA. These observations produced comparable radio data to the NVSS for the nearer clusters in both limiting radio luminosity and linear resolution. The details of these observations and results specific to the two clusters will be addressed in future papers.

\subsection{Spectroscopic Observations}

Long-slit spectra were obtained for many of the radio galaxies. In general, spectra were obtained for those galaxies lacking public velocity measurements and for galaxies whose far-infrared and radio fluxes did not provide an unambiguous classification as AGN or star forming \citep[see][hereafter called Paper II]{mil2001b}. Basic parameters of the observations -- telescopes, dates, spectrographs, and instrumental parameters -- may be found in Paper I. In general, a 2\arcsec{} slit was used in conjunction with a moderate resolution grating providing a resolution of $\sim 7\mbox{\AA}$ and spanning the entire optical wavelength regime. The large range in right ascension for the sample insured that nearly all observations were performed at low airmass, and consequently no slit rotation was attempted and each target galaxy was placed such that the slit was centered on the galaxy's nucleus. Observation of standard stars on each night were used to calibrate the spectra. The sensitivity functions derived from these standard star observations varied by only a few percent from night to night, with the largest variation occuring at the bluest edge of the spectrum where the system response dropped rapidly.

Extraction of the one-dimensional spectra was performed for two aperture sizes. First, a nuclear spectrum was extracted corresponding to the central 2\arcsec{} of each galaxy. A larger aperture of 15\arcsec{} was also extracted. The difference of these two apertures was used to assess the off-nuclear spectra of the galaxies. While the fixed angular size of the apertures means that slightly different linear scales were probed, using smaller apertures to explore the galaxies' nuclei would have been ineffective given the typical seeing during the observations and factors such as telescope tracking. For reference, 2\arcsec{} corresponds to 0.6--1.2 kpc for the 18 nearby clusters of the sample, and 2.0 kpc and 2.5 kpc for Abell 2256 and Abell 2255, respectively.

In addition to the long-slit data, the MX Spectrometer \citep{hill1986,hill1988} was used to obtain spectra in Abell 2255 and Abell 2256. This multifiber spectrograph is used in conjunction with the Steward Observatory 90'' telescope to gather spectra of up to 32 galaxies simultaneously. No attempt was made to flux calibrate these spectra, but analysis of their emission and absorption features was used to classify the target galaxies. The equivalent widths and spectral classifications thereof are included in this paper, and these galaxies were used in all subsequent analyses with the exception of the evaluation of dust extinction presented in Section \ref{sec:dust}. Details specific to these observations will be presented in a future paper.

\subsection{Spectroscopic Line Measurements}

The calibrated spectra were used to measure equivalent widths (EW) for a number of species. In an effort to introduce as little bias as possible, pre-defined rest-frame line and continuum regions were used (see Table \ref{tbl-EWdefs}). The EWs are calculated from the average fluxes in the defined line and continuum bands. The H$\alpha$--[NII] complex and [SII] doublet were deblended using the IRAF task SPLOT, with the continuum fit ``by eye.'' This also alleviated erroneous measurements caused by telluric absorption shifted into the pre-defined continuum bands. Visual inspection of the spectra during execution of SPLOT also served to identify lines contaminated by cosmic rays or telluric features. In addition to the line measurements, the strength of the 4000$\mbox{\AA}$ break \citep[$D_{4000}$, the ratio of average $F_\nu$ evaluated in the ranges $4050 - 4250 \mbox{\AA}$ to $3750 - 3950 \mbox{\AA}$;][]{bruz1983} was evaluated.

\placetable{tbl-EWdefs}

The resulting EWs represent the combination of contributions from stars and gas in the galaxies. As line ratio diagnostics apply specifically to emission lines arising from ionized gas, the measured EWs of the Balmer lines must be corrected to account for underlying stellar absorption. As the galaxies span a broad range in morphology, a standard correction was not applied (e.g., adding 1.5$\mbox{\AA}$ to the emission EW of each Balmer line). Instead, corrections based on the strength of the 4000$\mbox{\AA}$ break were applied, as this feature has been shown to be sensitive to star formation yet insensitive to factors such as metallicity \citep{drsh1987}. \citet{pogg1997} provide $D_{4000}$ values as a function of morphology in conjunction with the stellar contribution to H$\delta$ absorption. These data were fit linearly such that
\begin{equation}\label{eqn:d4_hd}
\mbox{EW}(\mbox{H}\delta)~=~-5.5D_{4000}~+~11.6
\end{equation}
for all $D_{4000}$ with a limit of EW(H$\delta$)=1.0 for higher values of $D_{4000}$. It was assumed that the H$\gamma$ and H$\beta$ lines have EWs equivalent to that of H$\delta$ \citep[e.g., see the results of][for continuous star formation at solar metallicity]{gonz1999} and that
\begin{equation}\label{eqn:ew_rel}
\mbox{EW}(\mbox{H}\alpha)~=~1.3~+~0.4\mbox{EW}(\mbox{H}\beta)
\end{equation}

\noindent
\citep{keel1983}. Thus, we evaluate $D_{4000}$ for each spectrum and translate it to the appropriate corrections for stellar Balmer absorption. The net correction at H$\alpha$ ranges from $1.7 - 3.8 \mbox{\AA}$. Table \ref{tbl-balcor} summarizes the corrections, and some discussion of their effects on calculated line ratios may be found in Section \ref{sec:discuss}. The Balmer-corrected EWs for all observed galaxies may be found in Tables \ref{tbl-EWs} and \ref{tbl-forbew}. The reported errors were determined from the rms variations in the line and continuum regions, plus an additional term to account for scatter in Equations \ref{eqn:d4_hd} and \ref{eqn:ew_rel} for the Balmer lines. In practice, these latter terms are usually much smaller than the variations in the line and continuum regions and affect the quoted errors for only the better measurements. Since the continuum regions may include real features in the spectra, the errors are likely upper limits to the true errors in the EW measurements. For reference, we have evaluated the S/N per pixel in the continuum region around H$\delta$ for the KPNO spectra. Among the cluster members, we obtain a mean S/N of 19 with a minimum of 8 and a maximum of 33. Analysis of a subsample of the APO spectra suggests these data are of slightly higher S/N (about $50\%$), although the improvement is less at longer wavelengths.

\placetable{tbl-balcor}
\placetable{tbl-EWs}
\placetable{tbl-forbew}

For the long-slit spectra, these corrected EWs were also used to calculate the line fluxes from the fluxes measured in the neighboring continuum. For this calculation, the error in the line flux was extrapolated from the error in the neighboring continuum, after accounting for the difference in size over which the line and continuum regions were defined (i.e., the line flux is generally measured over a smaller wavelength range than the continuum flux, thereby introducing additional uncertainty). Values for those galaxies with significant emission lines may be found in Tables \ref{tbl-LFs} for Balmer lines and \ref{tbl-forLF} for non-Balmer lines.

\placetable{tbl-LFs}
\placetable{tbl-forLF}

\subsection{Spectroscopic Classification of Galaxies}\label{sec:classes}

\subsubsection{Star Formation or AGN?}

Emission and absorption lines in the optical spectra were used to characterize the galaxies as dominated by star formation or an AGN. Diagnostics of the emission lines in spectra which contain them allow classification as HII-region like or AGN \citep[e.g.,][]{bald1981,veil1987,oste1989}. Within the AGN, the line ratios also allow characterization as either Seyferts or low-ionization nuclear emission-line region galaxies \citep[LINER;][]{heck1980}. For these purposes, the basic definitions of \citet{ho1996} were adopted: 
\begin{itemize}

\item{star forming or HII-region like:

[OI]$\lambda6300~<~0.08$H$\alpha$

[NII]$\lambda6584~<~0.6$H$\alpha$

[SII]$\lambda\lambda6717+6731~<~0.4$H$\alpha$} 

\item{LINER:

[OI]$\lambda6300~\geq~0.167$H$\alpha$

[NII]$\lambda6584~\geq~0.6$H$\alpha$

[SII]$\lambda\lambda6717+6731~\geq~0.4$H$\alpha$}

\item{Seyfert:

[OIII]$\lambda5007~\geq~3$H$\beta$

[OI]$\lambda6300~\geq~0.08$H$\alpha$

[NII]$\lambda6584~\geq~0.6$H$\alpha$

[SII]$\lambda\lambda6717+6731~\geq~0.4$H$\alpha$}

\end{itemize}

\begin{figure}
\figurenum{1}\label{fig-o3n2}
\epsscale{1.0}
\plotone{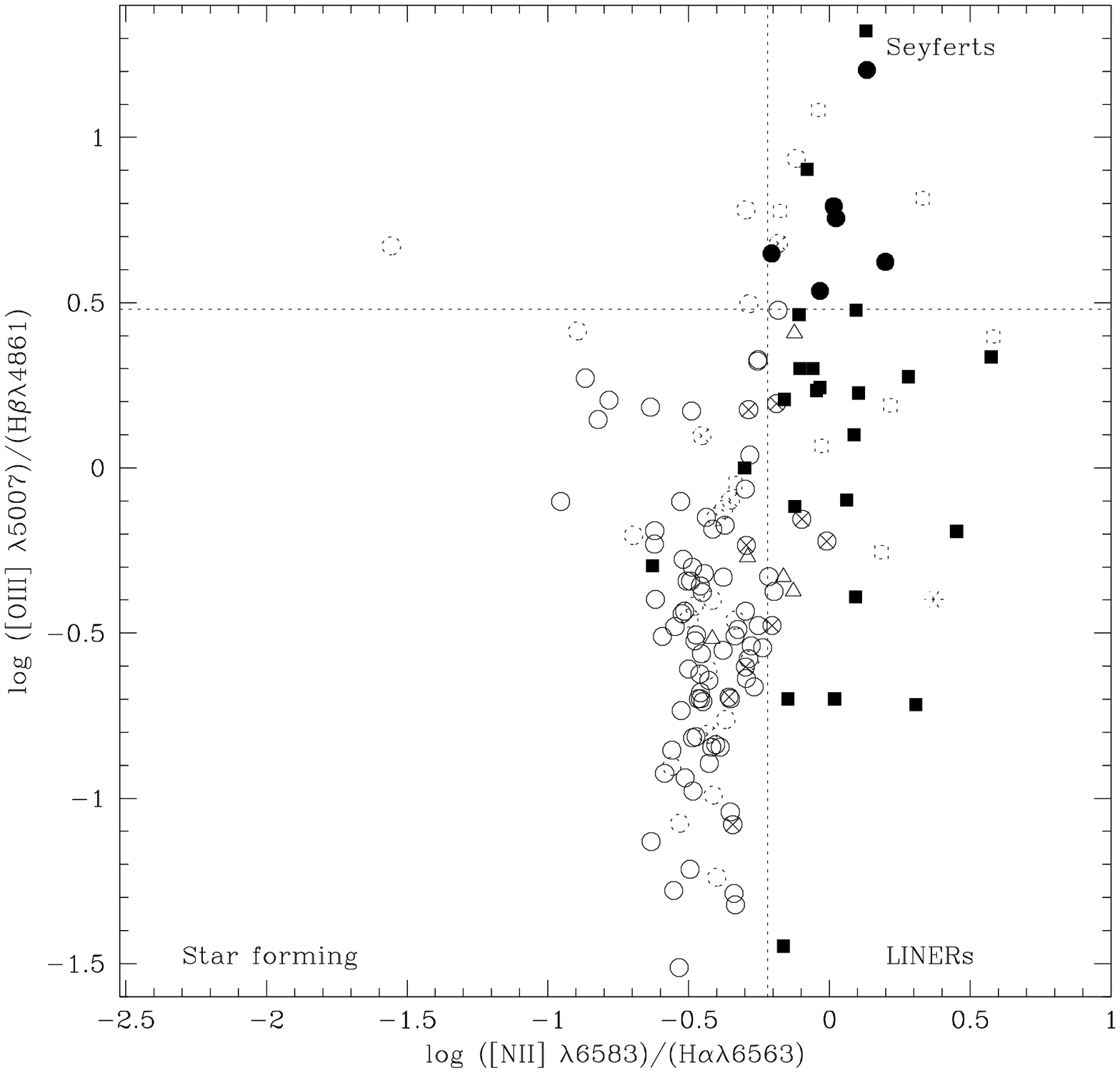}
\caption{Line ratio diagnostic diagram for [NII]$\lambda$6584 vs. [OIII]$\lambda$5007, each normalized by the nearest Balmer line to account for extinction. This diagram corresponds to the line measurements made for the central 2\arcsec{} of the galaxies. Open circles are star forming galaxies, filled squares are LINERs, filled circles are Seyferts, and open triangles are AGN with star formation noted in their off-nuclear spectra. Open circles with overlaid crosses are those galaxies for which the line ratio diagnostics did not produce significant results, and dotted symbols represent foreground/background galaxies.}
\end{figure}

\begin{figure}
\figurenum{2}\label{fig-o3s2}
\epsscale{1.0}
\plotone{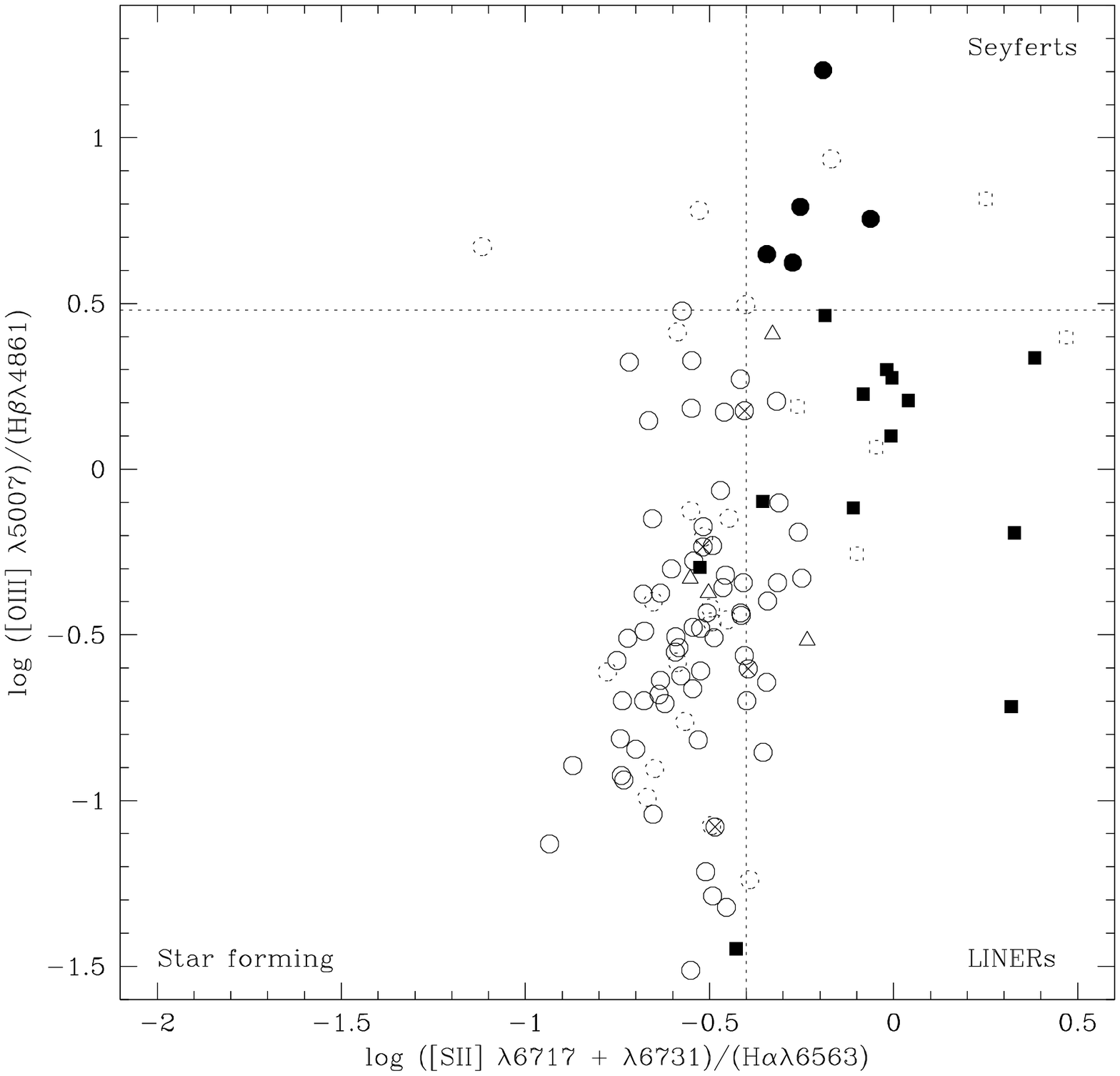}
\caption{Line ratio diagnostic diagram for [SII]$\lambda\lambda$6717+6731 vs. [OIII]$\lambda$5007, each normalized by the nearest Balmer line to account for extinction. The symbols and conventions are the same as those used in Figure \ref{fig-o3n2}.}
\end{figure}

\begin{figure}
\figurenum{3}\label{fig-o3o1}
\epsscale{1.0}
\plotone{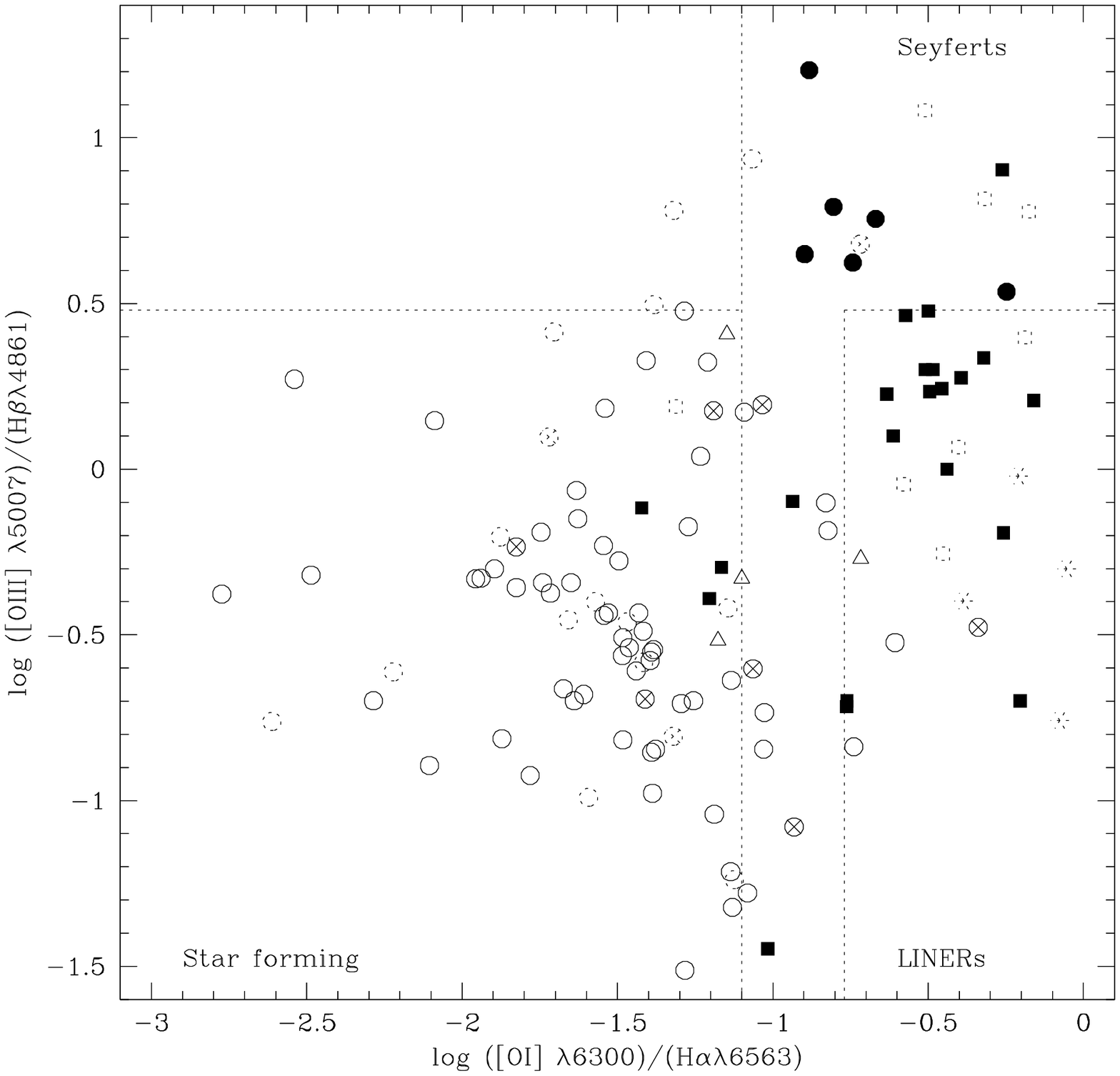}
\caption{Line ratio diagnostic diagram for [OI]$\lambda$6300 vs. [OIII]$\lambda$5007, each normalized by the nearest Balmer line to account for extinction. The symbols and conventions are the same as those used in Figure \ref{fig-o3n2}.}
\end{figure}

\begin{figure}
\figurenum{4}\label{fig-sample1}
\epsscale{1.0}
\plotone{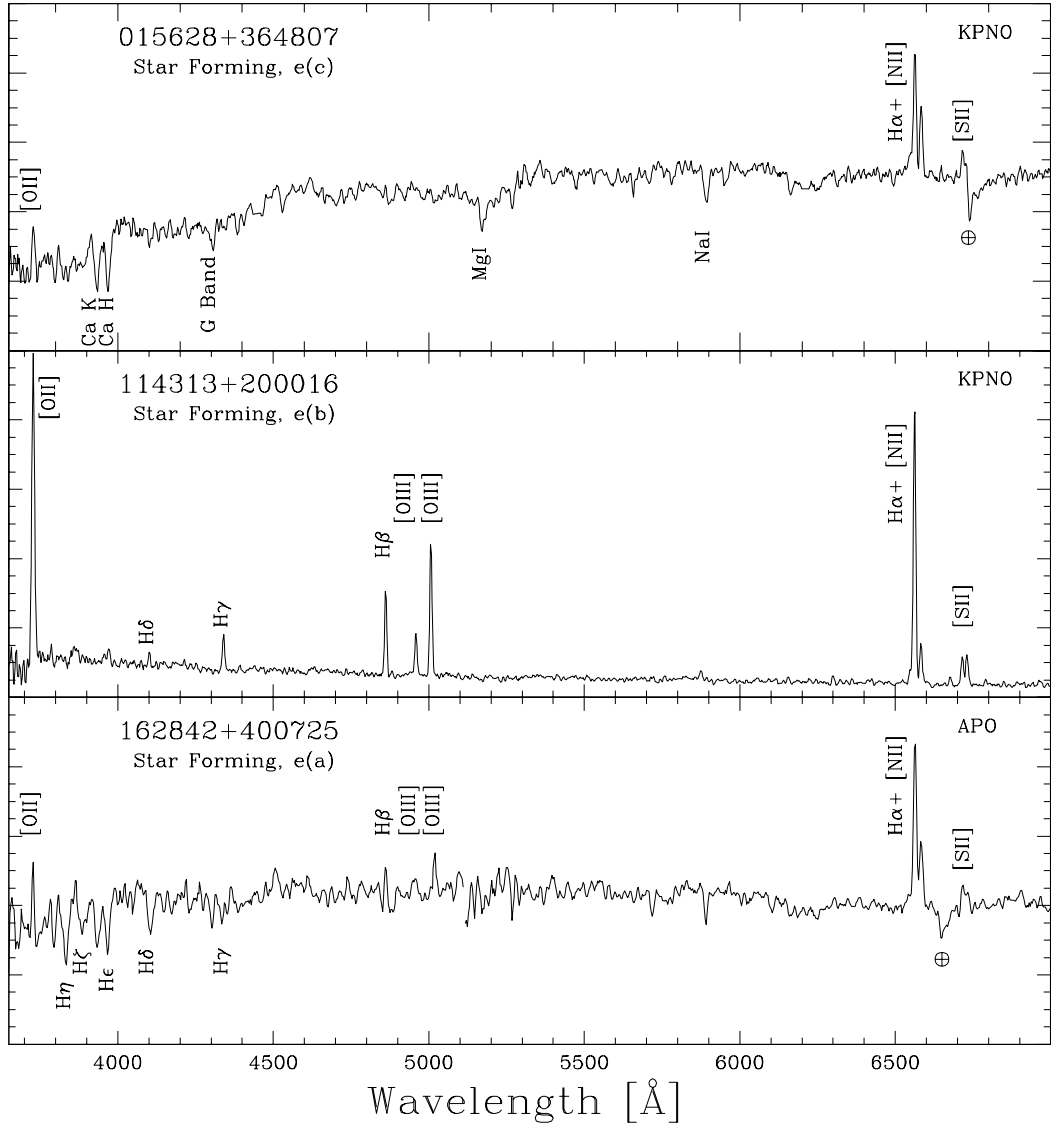}
\caption{Examples of galaxy spectra. The spectra have been smoothed to the system resolution, and the observatory where each was collected is found in the upper right corner. Emission and absorption lines are noted. Top: a normal star forming galaxy of type e(c); Middle: a star forming galaxy engaged in a burst of star formation, of type e(b); Bottom: a dusty star forming galaxy of type e(a).}
\end{figure}

\begin{figure}
\figurenum{5}\label{fig-sample2}
\epsscale{1.0}
\plotone{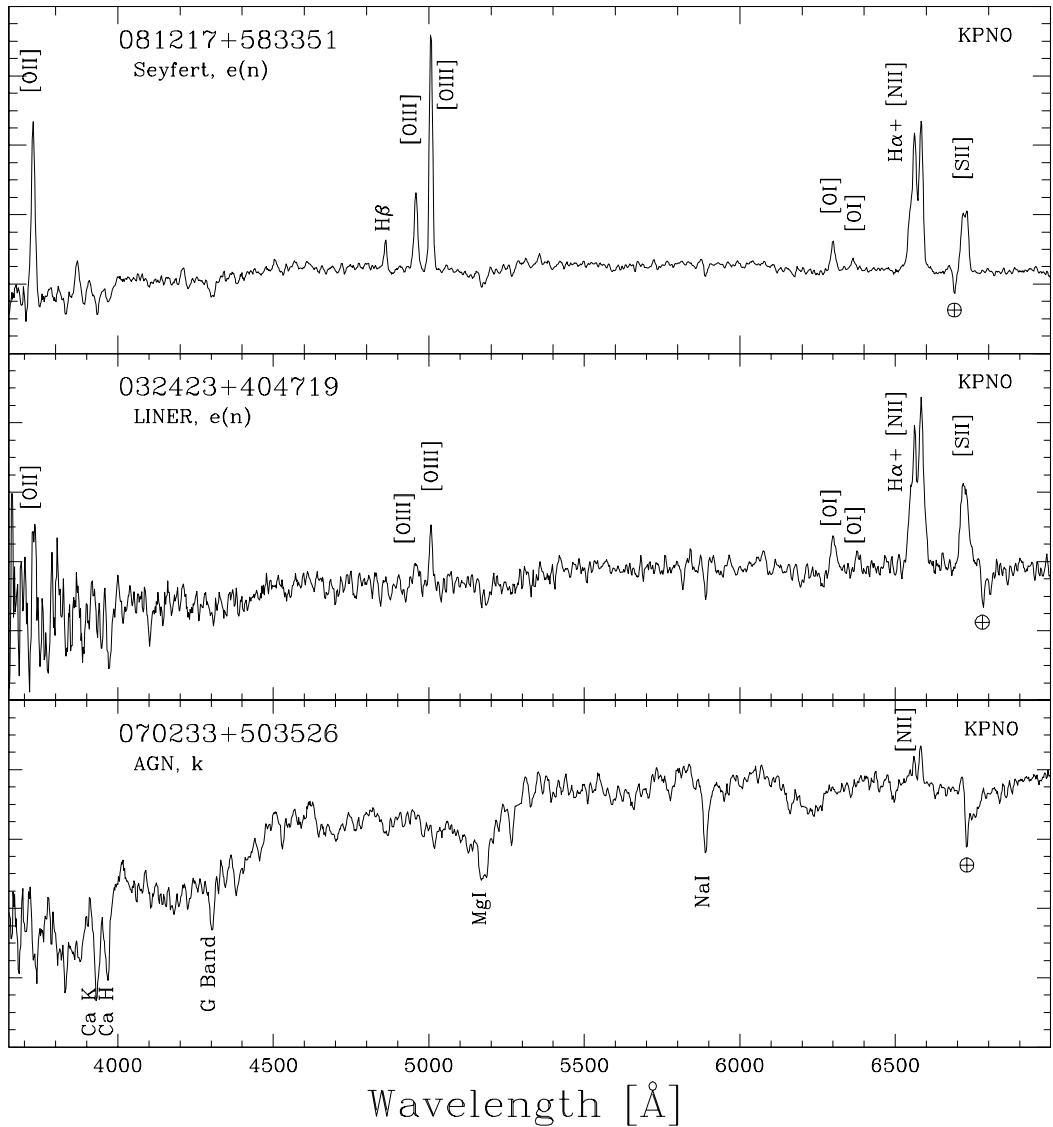}
\caption{More examples of galaxy spectra. Top: a Seyfert galaxy, which would be classified as e(n) in the MORPHS scheme; Middle: a LINER galaxy; Bottom: an AGN whose spectrum is largely that of an old stellar population, but slight emission of [NII] and [SII] (at edge of telluric absorption) are present.}
\end{figure}

\begin{figure}
\figurenum{6}\label{fig-sample3}
\epsscale{1.0}
\plotone{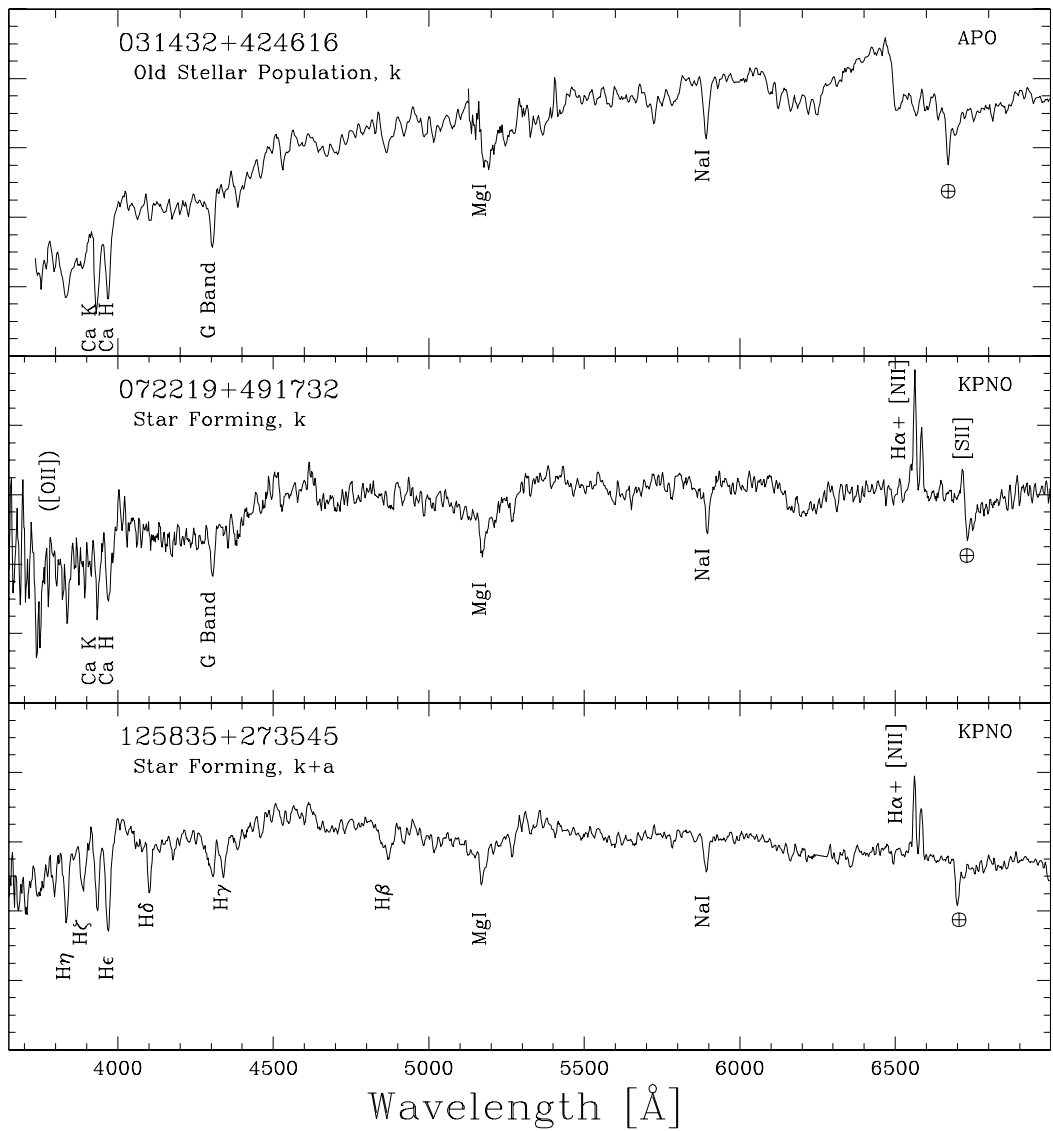}
\caption{More examples of galaxy spectra. Top: a galaxy whose spectrum is that of an old stellar population, or type k; Middle: a star forming galaxy which would be classified as k due to its lack of [OII] emission; Bottom: a star forming galaxy whose blue spectrum appears to be that of a k+a galaxy.}
\end{figure}

These tests are frequently displayed graphically as [NII]/H$\alpha$ vs. [OIII]/H$\beta$, [SII]/H$\alpha$ vs. [OIII]/H$\beta$, and [OI]/H$\alpha$ vs. [OIII]/H$\beta$. Using the EWs from Tables \ref{tbl-EWs} and \ref{tbl-forbew}, these three tests were performed. Figures \ref{fig-o3n2}--\ref{fig-o3o1} depict the results for the nuclear extraction aperture. Sample spectra covering each of the major classes of galaxies may be found in Figures \ref{fig-sample1}--\ref{fig-sample3}.

\placefigure{fig-o3n2}
\placefigure{fig-o3s2}
\placefigure{fig-o3o1}
\placefigure{fig-sample1}
\placefigure{fig-sample2}
\placefigure{fig-sample3}

Applying three separate tests naturally provides the possibility that not all three tests will agree. In part, this is the result of noise in the data which is simply represented by the error in the EW measurements. Using standard propagation of errors, the significance at which any single line ratio deviated from the dividing line between star forming galaxy and AGN was calculated. Deviations less than 1$\sigma$ in significance were ignored, which explained many of the contradictory test results. A further cause of error is misrepresentation of the true EWs for some lines. This was particularly true for the [SII] test, as the telluric feature at 6800$\mbox{\AA}$ is almost always redshifted into either the line or the defined continuum bands. In addition, bad columns and cosmic rays sometimes affected the spectra. For these reasons, the spectra were visually inspected to confirm that the line ratio results could be trusted and test results biased by such effects were ignored. To prevent the [SII] test from being rendered useless, whenever possible the deblended [SII] line EWs were summed and used instead of the combined value derived using the defined continuum and line regions. The reason for this choice was that in the deblending using the IRAF task {\it SPLOT}, an attempt to accurately fit the continuum independent of the redshifted absorption feature was made.

The spectra were first categorized by extraction aperture. In principle, the rules used were stringent in the classification of a galaxy as star forming. For each extraction aperture, the basic rules were that any single test producing a believable AGN result at the $1\sigma$ level or greater would classify the galaxy as an AGN. Within the AGN class, deference was given to LINERs. In practice, there were no galaxies with evidence for star formation which were classified AGN due to a single test producing a 1$\sigma$ AGN result. There were, however, a small number of galaxies (5) whose nuclear spectra possessed significant evidence for an AGN while their off-nuclear spectra were clearly those of more normal star forming galaxies. These were classified as ``mixed.''

Galaxies which did not contain any strong emission lines were considered to be AGN. These galaxies had easily distinguished old stellar populations with strong $D_{4000}$ and the usual absorption lines. The notion that even low radio luminosity galaxies whose spectra are dominated by an old stellar population are AGN has been noted in \citet{ho1999} among others. A number of galaxies had strong old stellar populations but slight noted emission of [NII]$\lambda6584$ and sometimes [SII]$\lambda\lambda6717+6731$ (see Figure \ref{fig-sample3}). This type of spectrum is fairly frequent in E and S0 galaxies; \citet{phil1986} estimate that $55-60\%$ of all such galaxies exhibit [NII] emission down to EWs of 0.5$\mbox{\AA}$. The same authors note that [NII] emission is more likely in radio-detected galaxies, and is a low-luminosity extension of the LINER class. Similarly, \citet{cozi1998} noted the presence of significant numbers of galaxies with these spectra in Hickson Compact Groups \citep[HCGs;][]{hick1982}. By subtracting a template old stellar population from these spectra, they ascertained that such objects were low-luminosity AGN, normally LINERs but in some instances very weak Seyfert 2s. To be classified as normal star forming galaxies, the [NII]-emitting galaxies in our sample would require a correction for stellar Balmer absorption greater than the largest correction assumed by Equations \ref{eqn:d4_hd} and \ref{eqn:ew_rel}. 

The assumption that {\it all} of the radio-detected galaxies with absorption line spectra correspond to AGN is probably not valid. The sample includes examples of galaxies with low radio luminosities and far-infrared fluxes suggestive of star formation, yet whose spectra do not include emission lines. One possible case for such a situation is star formation which occurs only in a small number of distinct HII regions that happen to fall outside the slit in our spectroscopy. A more extreme possibility is that star formation could be very heavily obscured by dust, thereby reducing the optical line emission to levels which do not stand out above the continuum in the spectra. These issues will be discussed further in Section \ref{sec:rlf}.

\subsubsection{``MORPHS'' Classification}

One of the primary goals of this work is to provide a local benchmark against which studies of clusters at higher redshift may be compared. For their studies of clusters at intermediate redshift, the MORPHS collaboration \citep[e.g.,][and subsequent papers]{dres1999,pogg1999} has applied an alternate system of galaxy classification. Driven largely by the inaccessibility of H$\alpha$ in higher redshift clusters, it is based on the strengths of the [OII] and H$\delta$ lines as indicators of current and recent star formation, respectively. The general division of this system is into passive and active galaxies,\footnote{This classification is based solely on optical spectra. In practice, all the galaxies of our sample are active as indicated by their radio emission. However, many radio galaxies have optical spectra which lack strong emission lines and are thereby called ``passive'' in the MORPHS terminology.} based on whether [OII] emission is present. Subdivision within these categories is based on the strength of H$\delta$ absorption. The categories are summarized in Table \ref{tbl-morphs}.

\placetable{tbl-morphs}

Classification based on line strengths is obviously affected by the procedure implemented in measuring the lines. We have attempted to reduce this effect by applying our line measurement procedure to a subset of the MORPHS spectra from \citet{dres1999}.\footnote{These spectra have kindly been made available for public FTP. Specifically, we used the spectra from the cluster Cl0939+47 for comparison.} This analysis determined that the EWs measured using our procedure were systematically weaker than those reported. Least squares fits to the data (including errors) produced the relations
\begin{mathletters}
\begin{equation}
\mbox{EW}(\mbox{H}\delta)_{M} = 1.27^{\pm0.05}\mbox{EW}(\mbox{H}\delta) - 0.64^{\pm0.18}
\end{equation}
\begin{equation}
\mbox{EW}(\mbox{[OII]})_{M} = 1.02^{\pm0.06}\mbox{EW}(\mbox{[OII]}) + 1.90^{\pm0.35}
\end{equation}
\end{mathletters}
The convention here is that lines seen in absorption have negative EWs. These corrections were applied in order to classify our cluster galaxies consistently with the MORPHS scheme. In practice, this changed the classification of only six of the cluster galaxies. Since the MORPHS sample is at much larger redshift than this sample, we have used the 15\arcsec{} aperture for these classifications as it corresponds more closely to the same physical size being sampled. 
\placetable{tbl-classes}

The instrument parameters used for the observations occasionally meant that the [OII] line was noisy or fell slightly off the spectrum. Cases for which the error in EW([OII]) was greater than its measured value have been classified as having no detectable [OII] emission. More common (about 25 instances) were cases in which the adopted blue continuum region for evaluating EW([OII]) was incomplete. These objects were visually inspected to ascertain whether [OII] was present, and have been classified based on this inspection. For 19 galaxies, [OII] fell off the measured spectrum and their MORPHS classification was based on other lines (star forming galaxies without [OII] were classified as ``active,'' galaxies without emission lines were classified as ``passive''). All classifications pursuant to these caveats are noted in Table \ref{tbl-classes}. 

It was also apparent from the spectroscopy that many of the galaxies which would be classified ``passive'' based on their lack of significant [OII] emission had significant H$\alpha$ emission. These could be recognized as star forming galaxies on the basis of their H$\alpha$, [NII], and [SII] lines. In fact, it is interesting to note that {\it all} of the cluster galaxies classified as k+a had detectable H$\alpha$ emission. In part, this result is to be expected given the nature of the sample. The radio selection picks out galaxies which either contain AGN or are actively forming stars. As a result, galaxies which are forming stars in highly dusty regions can lose their bluer emission lines to extinction, yet still possess the star formation signatures of H$\alpha$ and radio emission.

Classifications for the radio galaxies which were targets of optical spectroscopy observations are listed in Table \ref{tbl-classes}. These may be compared with Table 7 of Dressler et al.

\section{Analysis of Radio Galaxy Populations}\label{sec:analysis}

\subsection{Dust Extinction}\label{sec:dust}

The spectra provide additional avenues toward understanding the dusty star forming galaxies. Specifically, analysis of the Balmer decrements in the galaxies can indicate the amount of extinction and use of different sized extraction apertures can clarify its spatial distribution within the galaxies. Using the line flux measurements in Table \ref{tbl-LFs}, Balmer decrements were calculated for all star forming galaxies for which we obtained long-slit optical spectra. This was performed using both the nuclear extraction aperture (central 2\arcsec) and the off-nuclear aperture (15\arcsec{} aperture minus the central 2\arcsec{} aperture). Table \ref{tbl-balmer} summarizes the results.

The expected value of the Balmer decrement is 2.85, assuming Case B recomination, $T=10000$ K, and $n=10^4$ cm$^{-3}$ \citep{oste1989}. It can be seen that each studied classification of galaxy has a decrement in excess of this figure, which is not at all surprising. \citet{kenn92b} evaluated the Balmer decrements for a nearby sample of spiral galaxies, and found that a value of $\sim4$ was typical. This amounts to about one magnitude of extinction at H$\alpha$ \citep[assuming the Galactic Extinction law of][]{card1989}. In general, the Balmer decrements for the cluster radio galaxies in Table \ref{tbl-balmer} are consistent with this figure, with the average Balmer decrement for the full sample of spectroscopically-confirmed star forming galaxies being $4.04\pm0.13$ when measured over the 15\arcsec{} extraction aperture.

\placetable{tbl-balmer}

However, comparison of the nuclear and off-nuclear results demonstrates strong evidence for dusty nuclei in the cluster star forming galaxies. The average Balmer decrement measured from the nuclear extraction aperture was $4.78\pm0.16$, compared to $3.81\pm0.14$ for the off-nuclear extraction aperture. This difference is significant at the $4.6\sigma$ level. Analyzing the strength of the Balmer decrement as a function of spectral classification shows that the star forming galaxies which would be considered ``passive'' in the MORPHS scheme (i.e., those lacking significant [OII] emission) show the strongest evidence for this discrepancy. Their off-nuclear spectra appear quite normal, with an average decrement of $3.96\pm0.35$. But the nuclei of these galaxies are quite dusty, with an average decrement of $6.16\pm0.40$. This implies a visual extinction of two magnitudes, on average, in the nuclei of such galaxies and reaffirms the observed lack of [OII] in their spectra. Comparing these galaxies to the other star forming galaxies of the sample indicates that there is no statistical difference in their dust extinction outside the nuclei (i.e., in the galaxies' disks), but a highly significant ($>99.9\%$) difference in their nuclear dust extinction. No significant evidence is seen for dust extinction in the e(a) galaxies, which are presumably dusty starbursts. However, this is at least partially due to the small number of e(a) galaxies in the sample (9).

\subsection{Distributions}\label{sec:dist}

Due to the timescales for radio activity in galaxies, the distributions of radio galaxies in clusters provide a powerful diagnostic for the effect of environment. Based on both statistical and computational arguments, AGN are active radio emitters for $\sim 10^8$ years. The massive stars which power the cosmic ray electrons in star forming galaxies have lifetimes shorter than this figure, and synchrotron energy losses of these electrons limits the timescale over which they are strong emitters to a similar number. A representative velocity for the motion of a radio galaxy within a cluster is the cluster's velocity dispersion, which is around 1,000 km s$^{-1}$ for a rich cluster such as Coma. Given these figures, it is easy to calculate that the location of a radio galaxy in a cluster does not change greatly during its lifetime of radio activity. Thus, ignoring projection effects the locations of the radio galaxies are directly tied to their environments. Examination of a statistical sample of galaxies therefore serves as a powerful probe into the effect of the cluster environment on galaxy evolution.

Perhaps the most straightforward way to examine the distributions of various cluster populations is to fit them with King profiles \citep{king1966}. The actual function used by King is the representation of an isothermal profile in terms of spatial density. It may be integrated to describe the surface density, $\sigma (r)$, of galaxies within a given radius
\begin{equation}\label{eqn:king}
\sigma (r) = \pi r_c^2 \sigma_o \mbox{ln}[1 + (r/r_c)^2]
\end{equation}
where $\sigma_o$ is the central surface density and $r_c$ is the core radius. For a large sample of nearby clusters, \citet{adam1998} find $r_c \approx 170$ kpc for the cluster galaxies. Although fitting a spherically-symmetric function to the non-spherical distributions actually seen in clusters is imprecise, averaging over the 20 clusters of the sample greatly reduces the impact of this false assumption.

The radio galaxies of the sample were placed in radial bins of 0.25 Mpc and best-fitting core radii were calculated. To perform the fits, the total number of galaxies inside of 3 Mpc was used to normalize Equation \ref{eqn:king} for a given value of $r_c$. The $\chi ^2$ statistic was then computed for each value of $r_c$, with the value of $r_c$ producing the lowest $\chi ^2$ being adopted as the best-fit core radius. This procedure was applied for the complete sample as well as subsamples defined by activity and spectroscopic classification. Since star forming galaxies are more prevalent at lower radio luminosities and the full sample is based on a flux-limited catalog, the core radii fits were also performed for subsamples using a consistent radio luminosity cutoff. This was chosen to be the luminosity corresponding to a flux of 3.4 mJy in the farthest cluster for which the NVSS data was used, as the NVSS is over 99$\%$ complete at this flux \citep{cond1998}. A summary is provided in Table \ref{tbl-cores}.

\placetable{tbl-cores}

\begin{figure}
\figurenum{7}\label{fig-rcplot}
\epsscale{1.0}
\plotone{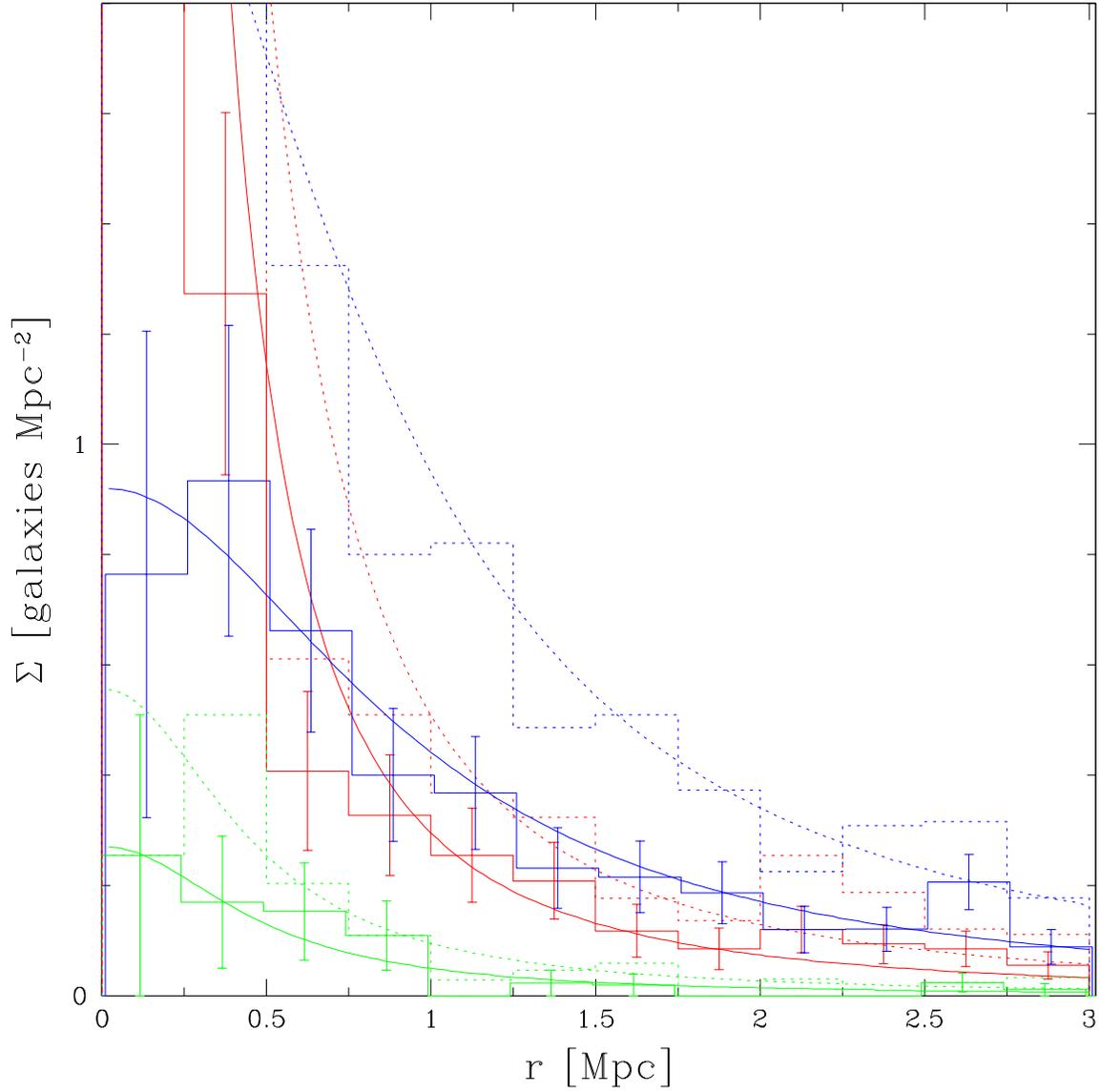}
\caption{Radial distributions of galaxies. The red data correspond to the AGN, the blue data to the star forming galaxies, and the green data to the dusty star forming galaxies. The solid histograms represent those galaxies for which the radio power was greater than $6.9\times10^{21}$ W Hz$^{-1}$, while the dotted histogram includes fainter galaxies. The corresponding best-fit king profiles are depicted in each case. The innermost bin for the AGN continues up to a value of 6.88 galaxies Mpc$^-2$; the figure has been truncated to better show the distributions of the star forming galaxies. The slight humps above 2 Mpc are caused by the overlap of A2197 and A2199.}
\end{figure}

The difference in the distributions of star forming galaxies and AGN is dramatic. Defining the AGN to include all emission-line AGN, galaxies with spectra dominated by old stellar populations, and galaxies presumed to be AGN on the basis of their far-infrared and radio fluxes (evaluated using $q$, the logarithm of the ratio of far-infrared to radio flux; see Section \ref{sec:rlf} for details), we find that the AGN are concentrated toward the cores of the clusters, with $r_c=180$ kpc ($r_c=130$ kpc for the radio luminosity-limited subsample). These values are consistent with the findings of \citet{adam1998} for cluster members in general. However, the core radius for the star forming galaxies (including those assignments made through optical spectroscopy and those based on $q$) is much greater, at $r_c=830$ kpc ($r_c=950$ kpc for the radio luminosity-limited subsample). Figure \ref{fig-rcplot} depicts the results.

\placefigure{fig-rcplot}

While issues of completeness may bias the results, the core radii of the distributions of specific spectroscopic classes of galaxies also prove interesting. Given the range of objectives for the observations (see Papers I and II), optical spectra were not obtained for all of the cluster radio galaxies. In general, galaxies with published redshifts and star forming/AGN classifications which could easily be made using radio and FIR data were not observed. Consequently, the core radii for specific classes must be viewed with caution. However, it is interesting to note that the core radius for the emission-line AGN population (Seyferts and LINERs) is larger than that seen for the AGN whose spectra resemble old stellar population objects. The core radii for the normal star forming galaxies, the e(c) class in \citet{dres1999}, are quite large (of order 2 Mpc). While this is certainly larger than the core radii found for the AGN populations, it is most likely biased by the large spatial extent used in our study. Previous studies of the clusters focused on the cluster cores so any star forming galaxies lacking redshifts were more likely to be located at large clustercentric radii.

Although the numbers are very small, we also see the same general trends noted by \citet{dres1999}. The passive galaxies are more centrally concentrated than the active galaxies, which seem to possess a decrease in numbers at the very cores of the clusters. The post-starburst galaxies are located more toward the cluster centers, but not as much as the passive galaxies.

In general, the picture that develops based on the radio-emitting galaxies is one in which the cores of the clusters are dominated by AGN and superposed on this strong peak in the galaxy distribution is a very broad hump of star forming galaxies. The influence of the clusters appears to be quite large, as evidenced by the large core radii for star forming galaxies and the presence of significant numbers of such galaxies well past the Abell radius. This segregation is also seen in the morphologies of galaxies \citep[e.g.,][]{dres1999} and the red and blue populations of cluster galaxies \citep[e.g.,][]{elli2001,morr1999,rako1997}.

And what of the dusty star forming galaxies, those which lacked emission lines in the blue portion of their spectra? In Table \ref{tbl-cores}, it can be seen that these galaxies appear nearer to the cluster cores than the star forming galaxies in general, but still more distant than the bulk of the AGN population. In fact, their distribution appears to peak at a radial distance of $\sim0.5$ Mpc and declines to near zero in the very centers of the clusters (see Figure \ref{fig-rcplot}). This important result argues that these dusty star forming galaxies appear to be caused by some cluster environmental effects.

These results may be tested statistically. The null hypothesis is that one class of galaxies is drawn from the same parent distribution as another. Kolmogorov-Smirnoff and Wilcox tests were applied to test this hypothesis and compare the radial positions (without binning) of various subpopulations. As would be expected, the distributions of the star forming galaxies and the AGN differed strongly. Using the full samples (including galaxies classified via $q$ values) both tests yielded less than a $0.2\%$ chance that the two classes were drawn from the same distribution. The results for the dusty star forming galaxies (i.e., those star forming galaxies classified as ``k'' or ``k+a'') were less conclusive. Comparing the dusty star forming galaxies to the rest of the star forming galaxies (again including those classified through $q$ values) produced a KS test result of a $93\%$ chance that the galaxies are drawn from different parent distributions. The Wilcox test, which effectively measures the significance at which the means of two distributions differ, produced a 1.4$\sigma$ ($84\%$) result. Thus, the dusty star forming galaxies appear to be more centrally-concentrated than the star forming galaxies in general, but the certainty of this conclusion is marginal.

\section{The Cluster RLF}\label{sec:rlf}

It is often desirable to classify galaxies as AGN or star forming on the basis of only their radio luminosity. Recent sub-millimeter surveys have discovered populations of high redshift galaxies whose optical counterparts are unidentified or extremely red. These galaxies are presumed to be dust-obscured starbursts, observable only at longer wavelengths. Because radio and far-infrared emission are so strongly correlated for nearby star forming galaxies, the use of radio as a star-formation indicator has increased in importance. Systems such as the VLA are capable of identifying high redshift objects at greater resolution and sensitivity than sub-millimeter observations, leading to their application in studies of the star formation history of the universe. However, the relative mix of AGN and star forming galaxies at lower radio luminosities is still the subject of debate.

The advent of the NVSS has enabled construction of the RLF from large samples using homogeneous sources of data. In particular, it has proven useful to marry the NVSS to large spectroscopic surveys and derive accurate RLFs from the resulting databases. \citet[][hereafter MG00]{mach2000} performed this exercise using the Las Campanas Redshift Survey \citep[LCRS;][]{shec1996}. Using 1157 LCRS galaxies matched with NVSS sources \citep[identified and classified in][]{mach1999}, they derived separate RLFs for star forming galaxies and AGN. \citet[][hereafter S02]{sadl2002} also derived RLFs for star forming galaxies and AGN, but used 662 galaxies from the 2dF Galaxy Redshift Survey \citep[2dFGRS;][]{coll2001}. The net RLFs determined by these two studies were extremely consistent. However, the relative composition of the RLFs differed at the lower radio luminosities, with the MG00 RLF indicating a turnover in the AGN RLF such that there were very few low-luminosity radio sources powered by AGN. The S02 RLF suggests that the AGN RLF does not exhibit a turnover, but continues to rise as one moves to lower radio luminosities (although star forming galaxies are still more common at these radio luminosities). The difference between the two may be the result of some bias in selection of the radio galaxies or in some subtle difference in the manner in which they are classified. In fact, S02 note that their sample includes 92 galaxies in common with MG00, of which $25\%$ are classified differently by the two studies. Of these, the majority are labelled star forming in MG00 and AGN in S02.

Our cluster sample provides an excellent basis to investigate the effects of the classification procedures used by MG00 and S02. Like each of those studies, we have large numbers of galaxies with known redshifts. We also possess the optical spectroscopy, magnitudes, radio luminosities, and far-infrared (FIR) data to apply each of their classification procedures and examine cases where discrepancies result. This analysis will help to elucidate whether there is a turnover in the RLF for AGN. Of course, our data certainly reflect a different selection than these two studies, so we can not perfectly separate classification differences from selection differences. This is simply because our data correspond to the RLF exclusively in clusters of galaxies, whereas the MG00 and S02 RLFs are more nearly field RLFs.\footnote{Actually, since the MG00 and S02 RLFs are evaluated from redshift surveys they include both cluster and field galaxies. They are consequently not strictly field RLFs and are better thought of as average RLFs.} Therefore, we are actually investigating two inter-related issues: first, whether our data can identify classification differences between the two studies and second, the importance of environment. This latter issue could also apply to the MG00 and S02 data, as their data may include different mixes of cluster and field radio galaxies.

The classification scheme used by S02 relies on the optical spectroscopy of the 2dFGRS, and is detailed in \citet{sadl1999}. It parallels our classfications discussed in Section \ref{sec:classes}, although the line ratio tests are only performed by eye. They place galaxies with an absorption-line spectrum in the class of AGN, as we have also suggested in Section \ref{sec:classes}.

\begin{figure}
\figurenum{8}\label{fig-RLF}
\epsscale{1.0}
\plottwo{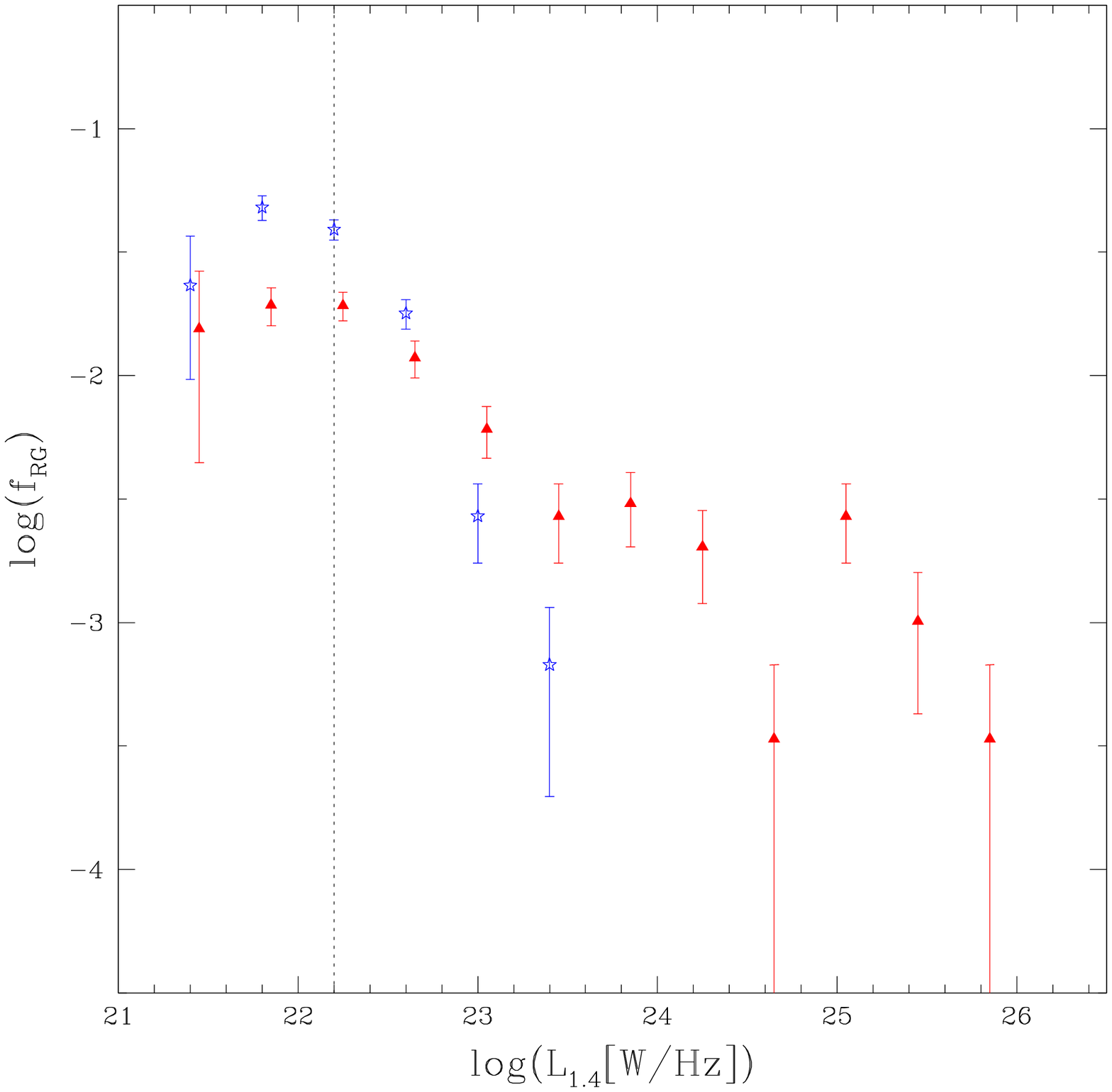}{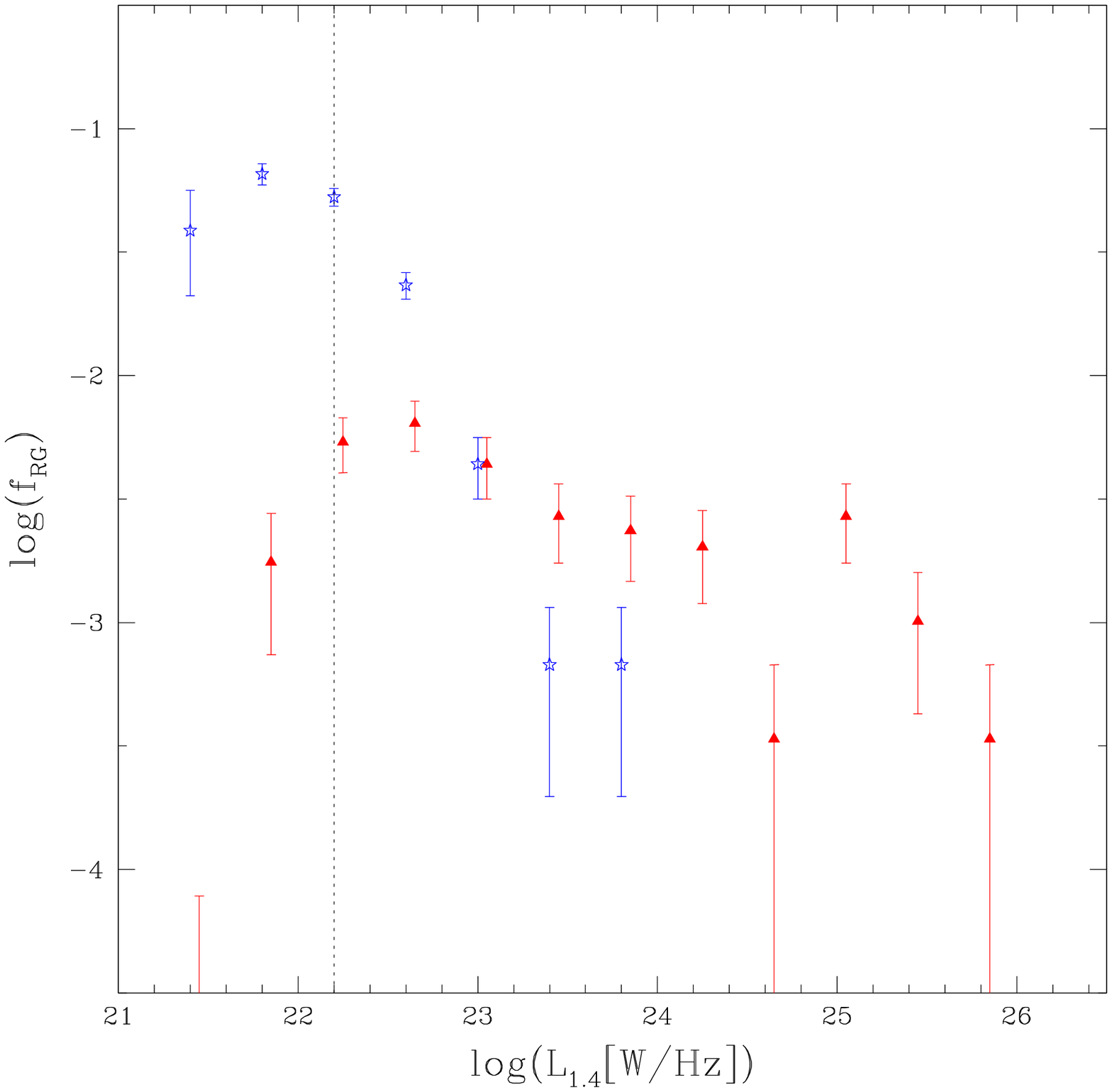}
\caption{The 1.4 GHz radio luminosity functions derived from the cluster data. At left, the classifications as discussed in Section \ref{sec:classes} were used and therefore are analagous to that presented in S02. At left, the classification scheme used by MG00 was adopted. AGN are represented by red filled triangles, whereas star forming galaxies are represented by blue stars. Note that these determinations have used $H_o=50$ km sec$^{-1}$ Mpc$^{-1}$ for consistency with MG00 and S02.}
\end{figure}

\placefigure{fig-RLF}

Our cluster RLF, constructed analogously to that of S02, is depicted on the left side in Figure \ref{fig-RLF}. We have used the same bin size and cosmological parameters as both S02 and MG00 ($H_o = 50$ km s$^{-1}$ Mpc$^{-1}$ and $q_o = 0.5$), but we have expressed the RLF as the fraction of all cluster galaxies (including radio non-detections) with radio luminosities in a given interval. Thus, it includes an implicit optical magnitude limit of $M_R=-20$ as described in Paper I and a correction for background counts. The correction for completeness was done in a simplistic manner. As before, only clusters for which the NVSS radio luminosity corresponded to a flux of 3.4 mJy or greater were used. This explains the apparently high lower luminosity limit in the figure, but does not affect the conclusions. 

MG00 classify radio galaxies based on three parameters: the radio source morphology and polarization, the FIR to radio flux ratio, and the radio to optical flux ratio. The fairly low angular resolution of the NVSS (45\arcsec) makes the first of these applicable mainly to powerful radio galaxies with large jets and lobes of radio emission, and as such is a nearly flawless way to determine if a radio source is powered by an AGN. The second relies on the strong correlation between FIR flux and radio emission observed for star forming galaxies, frequently parametrized by 
\begin{equation}
q~\equiv~\log \left( \frac{\mbox{FIR}}{3.75\times 10^{12}\mbox{W m}^{-2}} \right) - \log \left( \frac{S_{1.4GHz}}{\mbox{W m}^{-2}~\mbox{Hz}^{-1}} \right)
\end{equation}
where FIR is defined as
\begin{equation}
\left( \frac{\mbox{FIR}}{\mbox{W m}^{-2}} \right) ~\equiv~ 1.26 \times 10^{-14} \left( \frac{2.58 S_{60\mu \mbox{m}}+S_{100\mu \mbox{m}}}{\mbox{Jy}} \right)
\end{equation}
\citep{helo1985}. Many studies \citep[e.g.,][]{yun2001} have shown that the distribution of $q$ is very narrow, with $\langle q \rangle \sim 2.3$ and $\sigma \sim 0.2$. Hence, the IRAS 60$\mu$m and 100$\mu$m flux densities are used in conjunction with the NVSS radio flux density to evaluate $q$ and determine the likely nature of the radio source. For those radio galaxies with detected IRAS fluxes, MG00 adopt $q>1.8$ as star forming galaxies. The third parameter used by MG00 is the radio to optical flux ratio, defined as
\begin{equation}
r = \log (S_{1.4GHz}/f_R )
\end{equation}
where $f_R$ is calculated from the $R$ band magnitude:
\begin{equation}
f_R = 2.78 \times 10^{6 - 0.4R}.
\end{equation}
The MG00 galaxies presumed to be star forming (based on $q>1.8$) and AGN (based on extended radio emission) were shown to have $\langle r \rangle$ of $0.24 \pm 0.03$ and $1.72 \pm 0.04$ with $\sigma$ of 0.45 and 0.52, respectively. Thus, the large fraction of galaxies in the MG00 sample which lacked IRAS data were classified statistically using $r$.

On the right side of Figure \ref{fig-RLF} we plot our RLF derived using the same procedure as MG00. For the galaxies without FIR detections, we have adopted a value of $r=0.93$ as the division between AGN and star forming galaxies (chosen as the statistical midpoint between the two distributions, about 1.5$\sigma$ from each). 

\placefigure{fig-RLF}

It can be seen that the MG00 classification criteria result in a lower fraction of AGN contributing to the RLF at fainter luminosities than the S02 classification scheme. The apparent difference in the two RLFs is likely larger than Figure \ref{fig-RLF} would suggest, as our spectroscopy database includes only about half of all the cluster radio galaxies. The remainder were confirmed as cluster members using publicly-available velocities, and their classifications were based on the FIR-radio correlation. 

\begin{figure}
\figurenum{9}\label{fig-qandr}
\epsscale{1.0}
\plotone{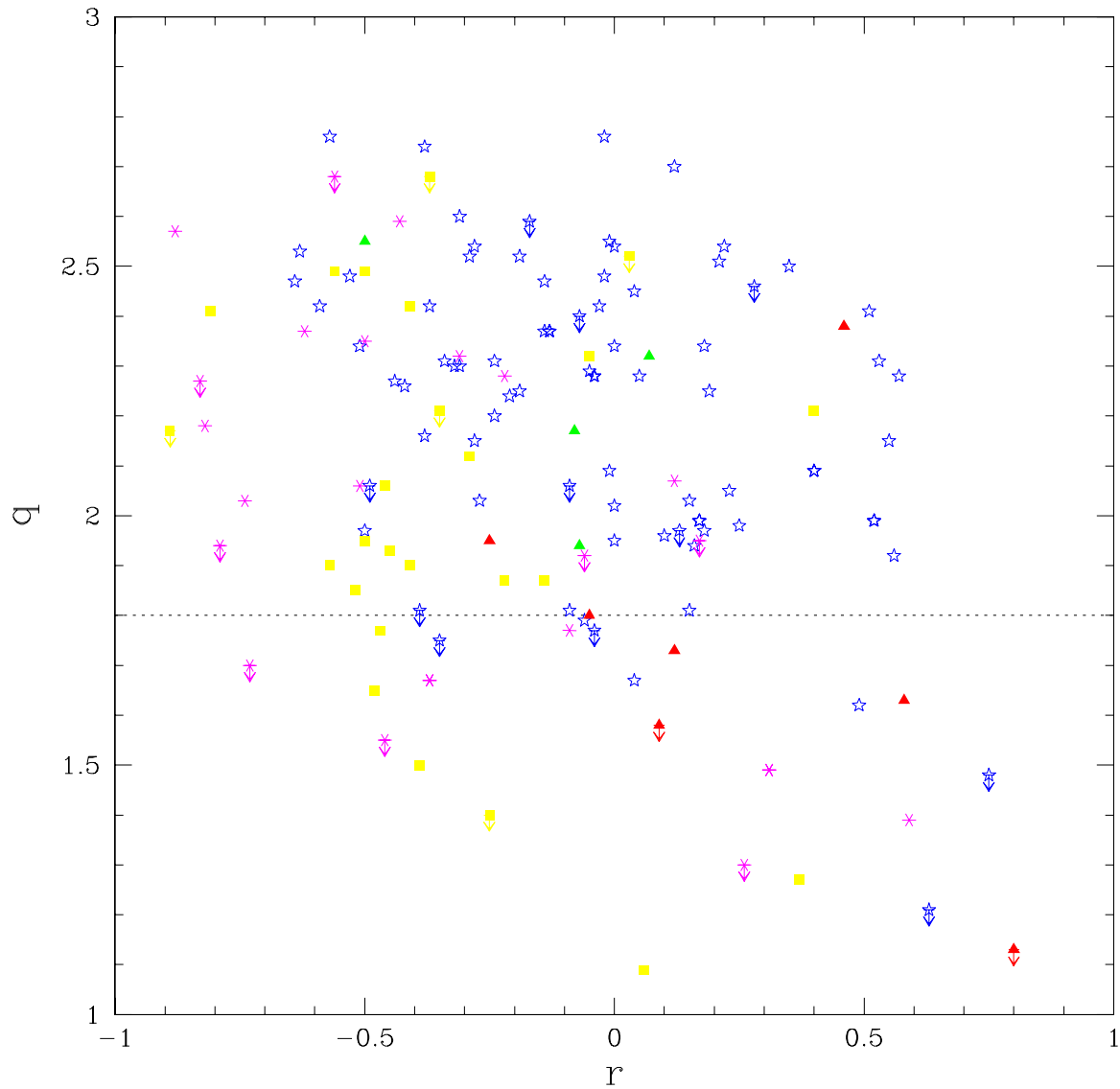}
\caption{Radio to optical flux ratio, $r$, vs. far-infrared to radio parameter, $q$. The symbols represent different types of spectra: blue stars, star forming galaxies; red filled triangles, emission-line AGN; magenta asterisks, pure absorption-line spectra; yellow filled squares, absorption-line spectra with weak emission of [NII] and [SII]. There are several points off to the lower right of the plot; each of these are powerful radio galaxies whose spectra are dominated by absorption lines.}
\end{figure}

\placefigure{fig-qandr}

Which galaxies are the source of this difference? In Figure \ref{fig-qandr}, we illustrate the differences in classifications by plotting $q$ and $r$ by type of optical spectrum. The data correspond to only those galaxies with FIR detections from IRAS, and in the event that the detection was only in one of the two bands the $q$ value is plotted as an upper limit calculated using three times the rms noise as the flux in the undetected band. The $r$ values covered in the figure show that in the absence of FIR data, all of these galaxies would be classified as star forming. In comparison with MG00, we find $\langle r \rangle = -0.16\pm0.02$ with a dispersion of 0.35. There are several additional points off the bottom right of the plot, corresponding to unambiguous AGN. The figure shows that a small number of emission-line AGN have strong enough FIR emission to classify them as star forming based on their values of $q$ and $r$. Presumably, this is due in part to star formation concurrent with the active nucleus in these galaxies; in fact, for four of these galaxies we identify an AGN in their nuclear spectra and star formation in their off-nuclear spectra. There are also a small number of star forming galaxies with unusually low $q$ values. These appear to be the result of cluster environmental effects, and are discussed in detail in Paper II.

More troubling are the large number of galaxies with absorption line spectra whose radio, optical, and FIR properties would suggest they are powered by star formation. These galaxies are AGN in S02, yet would be considered star forming in MG00. As our optical spectroscopy is not complete, there are likely significantly more such galaxies than are indicated in the figure. Clearly, it is important to better understand this population and the source of their radio emission.

It might be expected that the optical morphologies of these galaxies would shed light on the subject. A simple picture would have the elliptical galaxies as AGN and the spirals as star forming. Thus, we would expect the ellipticals to be either undetected in the FIR or have low enough $q$ values that they would be AGN in either classification scheme. The spirals would then be FIR detections and lie on the FIR-radio correlation. In this picture, the S0 galaxies would likely include a mix of AGN and star forming galaxies. 

To investigate this scenario, we obtained morphologies from NED (when available) for the galaxies with absorption line spectra. The galaxies were divided up by morphology into three groups: ellipticals, including galaxies classified as ``E/S0''; S0s, including barred S0's and galaxies classified as ``S0/a''; and later type galaxies. The later type galaxies in the sample were of type Sb and earlier (i.e., no Sc, Sd, or irregular galaxies), although some had noted rings. As the above hypothesis suggests, the majority of the later type galaxies were FIR sources (17 of 22). Of the remaining five later type galaxies with absorption line spectra, two were detected at 60$\mu$m but not at 100$\mu$m. Additionally, two of the three undetected galaxies were classified ``S?'' in NED and could easily be S0 galaxies. Some ellipticals and S0s were also FIR detections, although at lesser frequency than the later type galaxies (19 of 47, with about equal fractions for each the ellipticals and S0s). The difference is likely much larger if we account for bias in the observed galaxies. As described in Papers I and II, our targets for optical spectroscopy were based on $q$ values; galaxies without FIR detections and consequently low upper limits to $q$ (generally $q<1$) were usually not observed as they are almost certainly AGN.

While this can be interpretted as evidence for star formation in the later type galaxies, the lack of emission lines in their spectra must still be explained. It is likely that for some of these cases the star formation simply occurs in regions that lie outside the regions covered by the slit in the spectroscopic observations. The relative proximity of the sample translates to galaxy angular sizes around 1\arcmin, and our 15\arcsec{} aperture has been extracted from the center of the galaxies. The resulting spectrum for the Sa and Sb galaxies included in the later type category would then be dominated by the galaxies' bulge component. This is also especially true for the ring galaxies, in which the star formation would be expected to occur more in the ring. Any star formation outside the nuclear regions would be missed in the spectral analysis, yet the lower resolution of the radio and FIR data would include its effects. Related evidence is the absolute magnitudes of the later type galaxies with absorption line spectra, which are among the brighter galaxies of the sample ($M_R\lesssim-21$). This implies larger sizes and consequently a greater potential for missing emission lines due to aperture effects. The S02 data would suffer from the same bias, as the fibers used in the collection of the 2dFGRS spectra subtend approximately 2\arcsec{} on the sky.

An alternate explanation for the lack of emission lines in later type galaxies whose FIR and radio properties suggest star formation is that the star forming regions could be heavily obscured by dust. Such galaxies would be more extreme examples of the dusty star forming population noted in Section \ref{sec:analysis}, in which the dust extinction was severe enough to remove not only the bluer emission lines but H$\alpha$ as well.

Interestingly, S02 note that none of their galaxies with absorption line spectra fall on the FIR-radio correlation. Given the large number of absorption line spectra galaxies in our sample which do appear to lie on the FIR-radio correlation, it is natural to consider the importance of environment. It is possible that such galaxies are a cluster population, and the regions explored by S02 do not contain any clusters of galaxies at near enough redshift to be detected by the IRAS survey. Differing contributions of cluster and field galaxies may also give rise to variations in the RLFs of S02 and MG00.

Qualitatively, a noticeable difference between the cluster and general RLF can be seen in the crossover point where the net contributions due to star forming galaxies and AGN are equivalent. S02 quote a radio luminosity of $10^{23.2}$ W Hz$^{-1}$ for this point, whereas it can be seen to lie around $10^{22.7}$ W Hz$^{-1}$ in Figure \ref{fig-RLF}. Thus, the relative contribution of star forming galaxies to the RLF is lower in clusters than it is in the general local RLF. This result is not surprising, as it has been known that in general star formation is suppressed in the cluster environment \citep[e.g.,][]{balo1998,hash1998}. It can also be related to the morphology-density relationship, as clusters contain a larger fraction of elliptical galaxies and radio emission from these types is almost always associated with AGN rather than star formation. This latter effect is better seen by comparing the MG00 RLF with that of the right side of Figure \ref{fig-RLF}. Our RLF for the AGN population, constructed in the same manner as that of MG00, does not exhibit an obvious turnover. Thus, even though their relative contribution to the total RLF at low luminosities is much lower than that of star forming galaxies, it continues to be important.

Thus, the difference in the RLFs constructed by MG00 and S02 is at least partially due to classification. The population that seems responsible for this is the galaxies with spectra dominated by old stellar populations, and despite their lack of emission lines (including H$\alpha$) these galaxies are often also weak FIR emitters. In both the present study and that of S02 these galaxies are considered AGN, although this classification is tenuous. It is likely that a number of these galaxies are weak radio and FIR sources due to star formation, but any optical line emission is lost due to aperture effects, dust extinction, or some combination thereof. This would result in a mix of AGN and star forming galaxies somewhere between those presented in MG00 and S02.

\section{Discussion}\label{sec:discuss}

The full optical wavelength coverage which is easily available for the nearby clusters of the sample has demonstrated that the assumed activity class of galaxies is wavelength dependent. A fairly large number of the star forming galaxies in the sample would have been mistakenly classified as passive had their spectra not included the region around H$\alpha$. In these galaxies, extinction reduces the [OII]$\lambda3727$ line to a value on par with the noise. This type of spectrum was found in $\sim20\%$ of the star forming galaxies in the nearby clusters. Such objects would be missed in studies lacking adequate spectral coverage. A more definitive percentage can not be claimed here, as we lack spectroscopy of all the identified radio galaxies in the nearby clusters. Depending on their frequency among the galaxies assumed to be star forming on the basis of their far-infrared and radio fluxes, the fraction of dusty star forming galaxies may be greater or lower than this figure. An interesting subclass of these objects are the radio-identified k+a galaxies. The k+a class is believed to be post-starburst, yet {\em all} of the apparent k+a galaxies in the nearby clusters have H$\alpha$ emission indicating that they are presently forming stars. A possible parallel to these objects may be found in the radio-detected k+a galaxies in the intermediate redshift cluster Cl 0939+47 \citep{smai1999}. These distant k+a galaxies have radio-determined SFRs about an order of magnitude greater than those found in our nearby sample, placing them in the range of starburst galaxies. However, the incidence of star formation activity in k+a galaxies is likely dependent on the stringency applied in defining the class. Of the 21 galaxies identified as the strongest examples of post-starburst spectra in the LCRS, sensitive radio observations have demonstrated that the majority have very low upper limits on their SFRs \citep{mil2001c}. Even within this sample, though, examples of star forming galaxies with post-starburst spectra were identified.

\begin{figure}
\figurenum{10}\label{fig-d4hist}
\epsscale{1.0}
\plotone{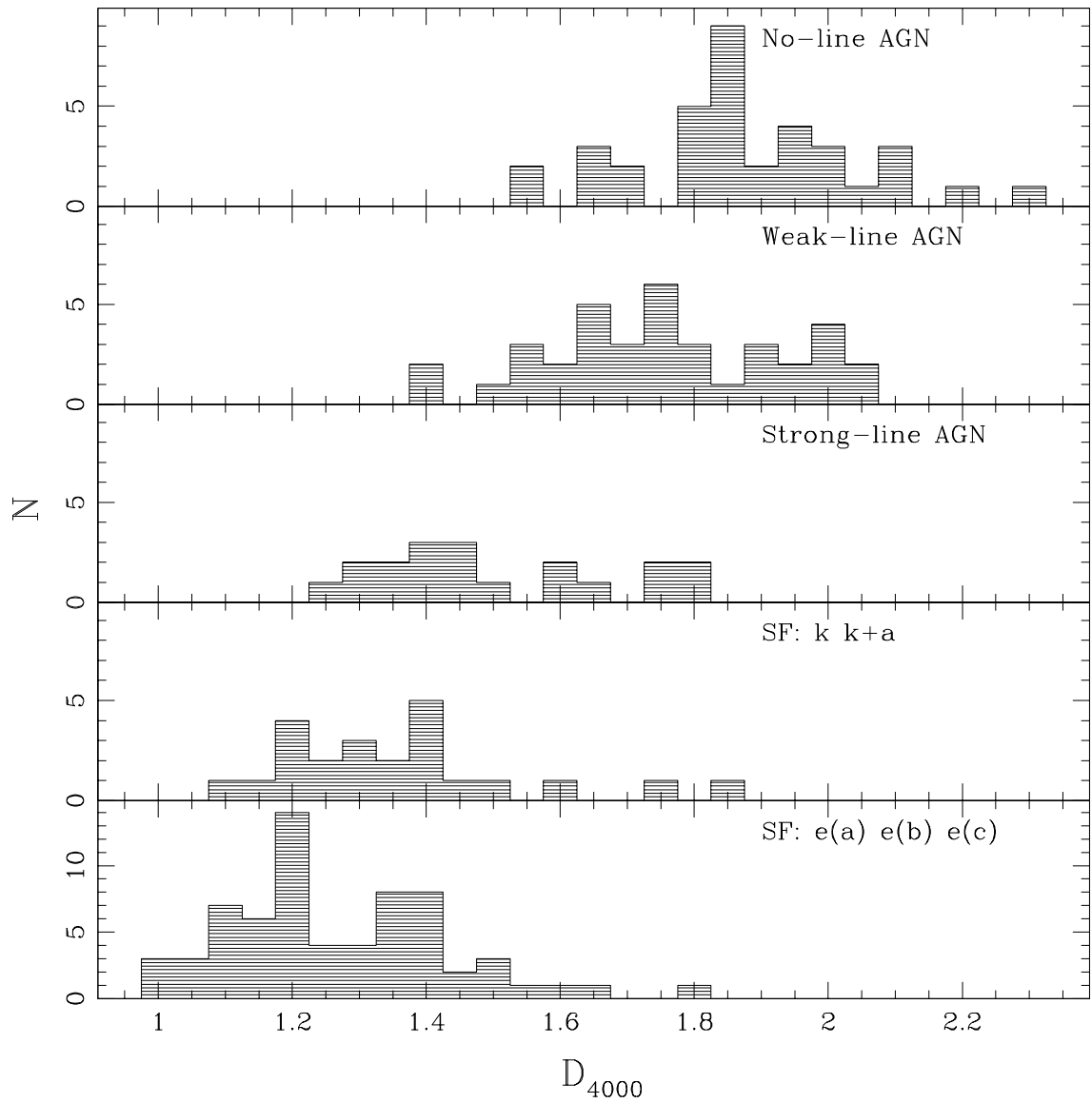}
\caption{Histograms of $4000\mbox{\AA}$ strength ($D_{4000}$) for cluster radio galaxies. See text for a description of the five classes.}
\end{figure}

Longer wavelength optical lines such as H$\alpha$ are difficult to observe in the intermediate redshift clusters of Butcher-Oemler studies. Given this, how can potential dusty star forming galaxies be identified in such systems? In Figure \ref{fig-d4hist} we show the $4000\mbox{\AA}$ break strengths ($D_{4000}$) of the cluster radio galaxies, broken down into five classes of spectra: 1) normal star forming galaxies, consisting of star forming galaxies of MORPHS class e(a), e(b), and e(c); 2) dusty star forming galaxies, consisting of star forming galaxies which would be considered passive (k and K+a) in the MORPHS scheme; 3) emission-line AGN, including strong Seyferts and LINERs which would be classified as e(n) in the MORPHS scheme; 4) weak emission-line AGN, whose spectra are dominated by absorption lines but include weak emission of [NII] and [SII]; and 5) galaxies whose spectra show only absorption lines. It can be seen that these classes are separated in terms of their $D_{4000}$ values. This provides an accessible means of identifying the dusty star forming galaxies in spectra lacking H$\alpha$ and nearby lines: the $4000\mbox{\AA}$ break in these galaxies is weak, due to the presence of star formation. A KS test comparing the $D_{4000}$ values of the dusty star forming galaxies with those of the pure absorption-line spectra galaxies produces a significance of $2.16\times10^{-10}$, or an extremely low probability that the galaxies are drawn from the same parent population. Similarly, a Wilcox test indicates that the mean $D_{4000}$ value of the dusty star forming galaxies is less than that of the pure absorption-line galaxies at over 6$\sigma$. Test results comparing the dusty star forming galaxies with the weak emission-line galaxies produce comparable results. Thus, while there inevitably is overlap in the $D_{4000}$ values of the various populations, it is relatively simple to statistically determine the likely nature of individual galaxies on the basis of $D_{4000}$ and nearby lines (particularly the presence or absence of [OII]).

\placefigure{fig-d4hist}

The importance of the dusty star forming galaxies may be underscored by their distribution. Their more centrally-concentrated distribution than the star forming galaxies with normal spectra argues that these galaxies are the product of some cluster environmental effect. There are a number of models attempting to describe such effects, including galaxy-galaxy interactions \citep[including ``harassment'';][]{moor1998}, interaction of infalling galaxies with the cluster tidal potential \citep[e.g.,][]{henr1996}, interaction of infalling galaxies with the intracluster medium \citep[e.g.,][]{gunn1972,fuji1999,quil2000}, and rapidly varying tides in group galaxies accreted by clusters \citep{bekk1999}. Most of these models predict bursts of star formation followed by its rapid truncation, but others suggest a more quiescent reduction in star formation as galaxies' gaseous halos are lost to the intracluster medium \citep{balo2000}. 

It is difficult to use the galaxy distributions to differentiate between these models using data that have been collected from a sample of 20 heterogenous clusters. But the spectra of the galaxies can be compared with model predictions to provide some additional perspective. Models indicate that production of H$\delta$ absorption large enough for galaxies to be classified as k+a require a starburst followed by rapid truncation of star formation \citep[e.g.,][]{pogg1999,shio2002}. This suggests that the identified star forming k+a galaxies are forming stars at current rates significantly less than they were in the past. \citet{shio2002} also investigated the more quiescent case where spiral galaxies lost their gaseous halo due to some cluster environmental effect, but did not experience a preceding starburst. In this case, star formation proceeds in the galaxy (albeit at a slower rate) as the galaxy uses up its molecular gas. The resulting spectra are similar to the k-type star forming galaxies of our sample: they do not have unusually strong H$\delta$ absorption, and their H$\alpha$ and radio emission are consistent with fairly low to normal star formation rates. Thus, it seems likely that multiple environmental effects are necessary to explain the different populations of cluster star forming galaxies.

Despite the difficulty in applying the various galaxies' distributions to evolutionary models, it is tempting to compare the distribution of the dusty star forming galaxies to galaxy types noted in other studies. The fact that the dusty star forming galaxies are most numerous near, but not in, the cluster centers is consistent with the distribution of S0 galaxies in clusters. This could reveal a possible connection between these objects, in the sense that the dusty star forming galaxies may be precursors to S0 galaxies. The same general distribution was found for the k+a galaxies in Dressler et al.

The Balmer decrement results strengthen these arguments. The majority of galaxy evolutionary models predict that starbursts are induced in the nuclei of affected galaxies. Consistent with this, we have found that the Balmer decrements for most cluster star forming galaxies are not unusual. In all cases except the star forming galaxies lacking [OII] emission, the Balmer decrements imply extinctions of about one magnitude near H$\alpha$ for both nuclear and off-nuclear spectra. However, this is not the case for the Balmer decrements of the star forming galaxies lacking [OII] emission. While the disks of these galaxies appear normal (on the basis of their off-nuclear decrements), their nuclei indicate substantially more dust extinction. Thus, it is probable that these galaxies have had a nuclear starburst excited and this starburst has led to large amounts of nuclear dust. Unfortunately, the present radio data can not shed any light on this theory. The resolution of the NVSS is too low to identify a nuclear starburst and the non-thermal radio emission associated with star formation is expected to originate over several kpc beyond the star forming regions. The range in galaxy orientations coupled with the fixed N-S slit also prevent a rigorous optical analysis using the spectra.

Evidence for an increased frequency of nuclear star formation in cluster galaxies has also been noted in other studies. \citet{rose2001} studied ten early-type galaxies in nearby clusters. These early-type galaxies were known to be unusual in the sense that they showed evidence for current or recent star formation. Using both long-slit spectroscopy and imaging in B and R, they noted that these galaxies had significantly greater nuclear activity than a comparable field sample. Using an objective-prism survey, \citet{moss2000} also argued for an increased frequency of nuclear starbursts in the galaxies of nearby clusters.

It is, of course, critical to address whether the identification of dusty star forming galaxies is biased. For example, it is possible that these galaxies are fairly normal star forming galaxies but some observational bias has rendered their [OII] emission undetectable. There is mild support for this notion, as these galaxies are somewhat more common in the lower redshift clusters of the sample. This implies a shorter wavelength for the redshifted [OII], which in turn means the line is observed at a wavelength where the system response is lower than it is for higher redshift [OII] emission. This increase in noise could effectively cover the [OII] line. A similar effect could result if these galaxies were simply of lower surface brightness, thereby reducing the S/N of the spectra and consequently the detectability of the [OII] line.

The other information derived for these galaxies leads us away from these possibilities. Primarily, the lack of an [OII] line is confirmed by the Balmer decrement analysis.  Furthermore, should the dusty star forming galaxies be misclassified due to observational bias we would expect them to be distributed in the same manner as the remainder of the observed galaxies. Lastly, we found no statistical differences in the magnitudes, radio luminosities, $q$ values (FIR-radio correlation), and $r$ values (radio-to-optical flux ratio) between the k-type star forming galaxies and the e-type star forming galaxies.

We did, however, explore the possibility that the increased Balmer decrements resulted from our corrections for stellar Balmer absorption. Figure \ref{fig-d4hist} shows that the $D_{4000}$ values for the k and k+a galaxies are larger than those of the e(a), e(b), and e(c) types. We find the mean value  of $D_{4000}$ for the dusty star forming galaxies is 1.37, compared to 1.27 for the normal star forming galaxies. Equations \ref{eqn:d4_hd} and \ref{eqn:ew_rel} show that this will increase the calculated Balmer decrements for the dusty star forming galaxies, but only by a negligible amount. Except for galaxies dominated by old stellar populations, our assumed stellar absorption at H$\alpha$ is always less than that assumed at H$\beta$. Therefore, adopting a uniform value for the Balmer correction (i.e., both H$\alpha$ and H$\beta$ are corrected by the same amount) would increase the derived decrements. Similarly, should our procedure underestimate the true absorption at H$\alpha$ our derived Balmer decrements would be lower than the real values.

In summary, the dusty star forming galaxies appear to be highly important in evolutionary studies of cluster galaxies. They can easily be overlooked due to their lack of strong emission lines, and appear to exist in fairly large numbers in the clusters being investigated. Their radial distribution in clusters along with the apparent concentration of activity in their nuclei present convincing evidence that they are an evolving cluster population.

Our analysis of the RLF identified a population of low luminosity radio sources with absorption line spectra. Understanding the nature of these radio galaxies is one of the most important issues raised by this study. We have noted that aperture bias can result in the misclassification of large early-type spirals as AGN when in fact their radio emission is due to low levels of star formation spread throughout their disks. However, this does not explain all such galaxies. For example, the regions probed by a fixed aperture depend not only on the distances to the galaxies but also on their orientations. The noted aperture bias will be stronger in galaxies which are oriented more closely to face on. For galaxies with more edge on orientations a fixed aperture will include disk light, so aperture effects alone are less likely to account for the lack of emission lines in such spectra. If these galaxies are star forming galaxies and not AGN, their star formation must take place in regions of heavy dust extinction.

Additional observations might help to reveal whether these galaxies are star forming or AGN. Higher resolution radio observations could be used to look for more compact structures. These could resolve jets (thereby identifying the galaxies as containing AGN) or individual star forming regions. This latter possibility would require a large increase in sensitivity, as the bulk of the emission at 20 cm due to star formation is synchrotron which originates over fairly large areas. It is believed that the deaths of massive stars accelerate cosmic ray electrons which emit as they spiral along galaxy magnetic field lines, thus extending the emission associated with a single star forming region over several kpc. High resolution radio observations could easily miss this low surface brightness emission. However, about $10\%$ of the radio emission is thermal bremsstrahlung from the HII regions themselves \citep{cond1992}. The detection of these regions would therefore be another step toward classifying the galaxies as star forming rather than AGN. Of course, this order of magnitude decrease in flux combined with the challenges of observing with higher resolution makes such observations difficult. An alternative approach would be to target features in the near IR, such as Pa$\alpha$ or the PAH features. 

\section{Conclusions}\label{sec:conclude}

We have used a comprehensive sample of active galaxies in 20 nearby Abell clusters to assess galaxy evolution in the cluster environment. Our multiwavelength data have revealed a number of interesting results:

\begin{itemize}

\item{Classification of galaxies using only bluer spectral features can mask a population of cluster star forming galaxies. These galaxies exhibit dust extinctions great enough to greatly reduce or even remove the presence of [OII] from their spectra. These galaxies represent somewhere near $20\%$ of all star forming galaxies in nearby clusters.}

\item{The distributions of radio galaxies in nearby clusters appear to be segregated by activity. Normal star forming galaxies are broadly distributed in clusters, extending well past the classical Abell radius whereas AGN are centrally concentrated. Galaxies which are actively forming stars but show unusually strong dust extinction appear to have a different distribution. Their core radius is intermediate to those of the star forming galaxies and the AGN, and they seem to peak at $\sim0.5$ Mpc while avoiding the very cores of clusters. This argues for such galaxies being the consequence of cluster environmental effects.}

\item{Spatially, the dust extinction in the dusty star forming galaxies is nuclear. Their Balmer decrements are large when measured for nuclear apertures (2\arcsec), yet normal when measured off the nucleus. Consequently, they appear to be the result of nuclear starbursts excited by some aspect of the cluster environment.}

\item{In higher redshift clusters for which H$\alpha$ is beyond the range of the optical spectrum, candidates for these dusty star forming galaxies may be identified on the basis of their $D_{4000}$ values. The dusty star forming galaxies have signficantly lower $D_{4000}$ than non-star forming galaxies whose shorter wavelength spectra are characteristic of old stellar populations.}

\item{There is a population of galaxies with spectra dominated by old stellar populations whose radio luminosities place them in the realm of either star forming galaxies or low luminosity AGN. These galaxies are frequently weak FIR detections, and ratios of their radio, optical, and FIR fluxes do not unambiguously identify them as star forming or AGN. It is likely that the assumed classification of these galaxies is responsible for differences in the RLFs reported by different authors.}

\item{The radio luminosity at which AGN and star formation contribute equally to the overall RLF is greater in clusters than it is in large volume-limited samples. This effect can be traced to the lower fractions of star forming galaxies in clusters relative to the field.}

\end{itemize}

Even though these results have been obtained using a large spectroscopic database, additional spectroscopy would prove very useful. The fairly large fraction of galaxies classified on the basis of radio and FIR fluxes means that more precise estimates of the frequency of various spectral types (including dusty star forming galaxies, AGN, etc.) can not be made.

While these results have been obtained from the collective sample, additional information can be obtained by assessing each cluster individually in comparison to the rest of the sample. Specifically, are these evolutionary results more common in some clusters than others? If so, are they a function of factors such as richness or dynamical state? It is possible that an understanding of these issues will help to understand any underlying causes for discrepancies in evolutionary studies at higher redshift, if these causes can be related to selection effects in the different higher redshift studies.

\acknowledgments

NAM acknowledges the NRAO predoctoral program and a National Research Council Associateship Award for support of this research. The authors also thank John Hill for use of the MX Spectrometer to collect velocities of galaxies in Abell 2255 and Abell 2256, and Bill Oegerle and Rajib Ganguly for providing additional velocity measurements for galaxies in Abell 2255. Finally, the authors thank the anonymous referee for thoughtful comments which strengthened the paper.

\clearpage

\onecolumn

\clearpage

\begin{deluxetable}{l c c c}
\tablecolumns{4}
\tablecaption{Line and Continuum Regions for Spectroscopy\label{tbl-EWdefs}}
\tablewidth{0pt}
\tablenum{1}
\tablehead{
\colhead{Line Name} & \colhead{Continuum 1} & \colhead{Continuum 2} & \colhead{Line}}
\startdata
$[$OII$]~\lambda$3727             & 3653.0--3718.0 & 3738.0--3803.0 & 3718.0--3738.0 \\
H$\delta~\lambda$4102           & 4055.0--4080.0 & 4120.0--4145.0 & 4091.0--4111.0 \\
H$\gamma~\lambda$4340           & 4230.0--4270.0 & 4360.0--4400.0 & 4330.0--4350.0 \\
H$\beta~\lambda$4861            & 4786.0--4851.0 & 4871.0--4939.0 & 4851.0--4871.0 \\
$[$OIII$]~\lambda$4959            & 4874.0--4939.0 & 5017.0--5082.0 & 4949.0--4969.0 \\
$[$OIII$]~\lambda$5007            & 4874.0--4939.0 & 5017.0--5082.0 & 4997.0--5017.0 \\
$[$OI$]~\lambda$6300              & 6200.0--6260.0 & 6340.0--6400.0 & 6285.0--6315.0 \\
H$\alpha\lambda6563+[$NII$]\lambda\lambda$6548,6584 & 6460.0--6520.0 & 6610.0--6670.0 & 6535.0--6595.0 \\
$[$SII$]~\lambda\lambda$6717,6731 & 6640.0--6700.0 & 6750.0--6810.0 & 6705.0--6745.0 \\
\enddata

\end{deluxetable}

\begin{deluxetable}{l c c c c}
\tablecolumns{5}
\tablecaption{$D_{4000}$ and Balmer Absorption\label{tbl-balcor}}
\tablewidth{0pt}
\tablenum{2}
\tablehead{
\colhead{Type} & \colhead{$D_{4000}$} & \colhead{H$\delta$} & \colhead{H$\beta$} & \colhead{H$\alpha$}}
\startdata
E  & 2.21 & 0.9 & 0.9 & 1.7 \\
Sa & 1.91 & 1.2 & 1.2 & 1.8 \\
Sb & 1.75 & 2.0 & 2.0 & 2.1 \\
Sc & 1.54 & 3.2 & 3.2 & 2.6 \\
Sd & 1.42 & 3.7 & 3.7 & 2.8 \\
Ex & 1.31 & 4.1 & 4.1 & 2.9 \\
\enddata

\tablecomments{Type, $D_{4000}$, and H$\delta$ are reprinted from Poggianti \& Barbaro 1997. H$\beta$ is assumed to be equal to H$\delta$, and H$\alpha$ is calculated using the relationship presented in Keel 1983.}

\end{deluxetable}

\begin{deluxetable}{c c c c c c c c}
\tablecolumns{8}
\tablecaption{$D_{4000}$ and Balmer EWs for Cluster Radio Galaxies\tablenotemark{*}\label{tbl-EWs}}
\tabletypesize{\scriptsize}
\tablewidth{0pt}
\tablenum{3}
\tablehead{
\colhead{ID} & \colhead{Ap} & \colhead{$D_{4000}$} & \colhead{H$\delta$} & 
\colhead{H$\gamma$} & \colhead{H$\beta$} & \colhead{H$\alpha$+[NII]} & \colhead{H$\alpha$} \cr
\colhead{} & \colhead{} & \colhead{} & \colhead{$\lambda 4102$} & 
\colhead{$\lambda 4340$} & \colhead{$\lambda 4861$} & 
\colhead{$\lambda$6548,6563,6584} & \colhead{$\lambda 6563$}}
\startdata
015129+360356 & A & 1.34(0.08) & 0.8(1.9) & 3.6(1.4) & 6.1(1.0) & 34.0(2.7) & 26.6(2.7) \\*
              & N & 1.67(0.15) & -2.8(2.2) & 1.6(2.0) & 2.5(1.2) & 27.7(3.0) & 19.3(3.0) \\*
              & O & 1.29(0.08) & 1.5(2.0) & 3.9(1.5) & 6.8(1.1) & 35.7(3.0) & 27.7(3.0) \\
015240+361016 & A & 1.91(0.08) & 1.2(0.8) & 0.7(0.8) & -0.5(0.4) & 3.6(0.6) & \nodata \\*
              & N & 2.03(0.08) & 1.4(0.6) & 1.8(0.8) & -0.7(0.4) & 4.7(0.7) & \nodata \\*
              & O & 1.85(0.09) & 1.4(1.1) & 0.4(0.9) & -0.1(0.4) & 3.0(0.7) & \nodata \\
015246+360907 & A & 1.94(0.24) & -0.3(2.5) & 5.2(3.0) & -0.4(1.1) & 14.9(1.9) & 5.1(1.9) \\*
              & N & 1.92(0.29) & 1.0(3.3) & 5.2(4.0) & 2.6(1.7) & 27.1(2.5) & 8.1(2.5) \\*
              & O & 1.94(0.25) & 1.9(3.6) & 5.2(3.3) & -1.1(1.3) & 11.5(2.1) & 3.2(2.1) \\
015254+360311 & A & 1.38(0.08) & -0.8(1.7) & 1.6(1.8) & 6.6(1.5) & 42.2(2.9) & 33.5(2.9) \\*
              & N & 1.31(0.09) & 2.9(1.9) & 2.4(2.2) & 2.3(1.3) & 34.4(3.3) & 26.4(3.3) \\*
              & O & 1.40(0.09) & -1.8(1.9) & 1.4(2.0) & 7.9(1.9) & 44.5(3.4) & 34.7(3.4) \\
015628+364807 & A & 1.59(0.06) & 2.2(0.7) & 2.3(0.6) & 2.4(0.4) & 14.1(0.8) & 10.4(0.8) \\*
              & N & 1.65(0.06) & 1.4(0.8) & 1.6(0.6) & 2.2(0.4) & 14.4(0.8) & 10.8(0.8) \\*
              & O & 1.56(0.06) & 2.5(1.0) & 2.5(0.8) & 2.3(0.5) & 13.8(1.1) & 10.0(1.1) \\
\enddata

\tablenotetext{*}{The complete table will be available in the electronic version of the Astronomical Journal.}

\tablecomments{With the exception of H$\alpha$+[NII], these EWs have been corrected for the assumed Balmer absorption component caused by starlight in the galaxies. The adopted convention has positive EWs representing emission lines and negative EWs representing absorption, with errors placed in parentheses after the values. Where blends occur, the deblending has been performed in IRAF using the task {\it splot} as described in the text. The `Aperture' column denotes the extraction aperture, with `A' denoting the full 15\arcsec{} aperture, `N' denoting the 2\arcsec{} aperture, and `O' denoting the difference between the two. `M' denotes that the spectrum was obtained with the MX multifiber spectrometer, and no spatial information can be obtained. Such spectra are not flux calibrated, and consequently meaningful values of $D_{4000}$ can not be calculated. For this reason, no correction for Balmer absorption has been applied to these spectra. Values are not presented in two cases: 1) all or some portion of the defined continuum region lay off the CCD, and 2) blended lines that were too weak to be deblended.}

\end{deluxetable}

\begin{deluxetable}{c c c c c c c c c c}
\tablecolumns{10}
\tablecaption{Forbidden Line EWs for Cluster Radio Galaxies\tablenotemark{*}\label{tbl-forbew}}
\tabletypesize{\tiny}
\tablewidth{0pt}
\tablenum{4}
\tablehead{
\colhead{ID} & \colhead{Ap} & \colhead{[OII]} & \colhead{[OIII]} & \colhead{[OIII]} & 
\colhead{[OI]} & \colhead{[NII]} & \colhead{[SII]} & \colhead{[SII]} & \colhead{[SII]} \cr
\colhead{} & \colhead{} & \colhead{$\lambda 3727$} & \colhead{$\lambda 4959$} & 
\colhead{$\lambda 5007$} & \colhead{$\lambda 6300$} & \colhead{$\lambda 6584$} &
\colhead{}  & \colhead{$\lambda 6717$} & \colhead{$\lambda 6731$}}
\startdata
015129+360356 & A & 2.0(8.4) & 1.7(1.1) & 0.7(1.0) & 1.1(1.5) & 10.0(2.7) & 11.6(3.5) & 4.2(3.5) & 4.4(3.5) \\*
              & N & -2.3(13.2) & 1.1(1.5) & 0.5(1.5) & 0.1(1.9) & 6.6(3.0) & 9.0(3.0) & 3.3(3.0) & 4.4(3.0) \\*
              & O & 2.6(8.9) & 1.8(1.2) & 0.0(1.1) & 1.3(1.8) & 9.9(3.0) & 12.3(4.0) & 4.5(4.0) & 4.2(4.0) \\
015240+361016 & A & -1.8(4.0) & 0.6(0.4) & -0.4(0.4) & -0.1(0.5) & 0.6(0.6) & 0.2(0.7) & \nodata & \nodata \\*
              & N & -0.5(3.5) & 0.8(0.4) & -0.6(0.4) & 0.1(0.4) & 1.9(0.7) & 0.9(0.7) & \nodata & \nodata \\*
              & O & -1.8(5.1) & 0.5(0.5) & -0.3(0.5) & -0.3(0.6) & 0.6(0.7) & -0.3(0.8) & \nodata & \nodata \\
015246+360907 & A & 0.0(0.0) & 0.6(1.4) & 0.0(1.4) & 1.6(1.4) & 11.0(1.9) & 8.7(2.0) & 6.4(2.0) & \nodata \\*
              & N & 0.0(0.0) & 0.6(1.7) & 0.5(1.8) & 1.4(1.8) & 16.4(2.5) & 12.9(2.7) & 16.9(2.7) & \nodata \\*
              & O & 0.0(0.0) & 0.7(1.5) & -0.2(1.5) & 1.6(1.7) & 5.2(2.1) & 7.6(2.2) & 4.5(2.2) & \nodata \\
015254+360311 & A & 9.0(10.0) & 0.5(1.4) & 0.7(1.3) & 0.4(1.6) & 10.2(2.9) & 0.4(2.1) & 2.5(2.1) & \nodata \\*
              & N & 9.0(11.2) & -0.2(1.5) & -0.4(1.5) & -0.1(2.1) & 7.4(3.3) & -2.5(2.5) & \nodata & \nodata \\*
              & O & 10.1(11.5) & 0.7(1.6) & 1.0(1.5) & 0.6(2.0) & 11.7(3.4) & 1.3(2.5) & \nodata & \nodata \\
015628+364807 & A & 4.3(2.5) & 0.7(0.5) & 0.0(0.5) & 0.8(0.4) & 5.0(0.8) & 0.7(0.6) & 1.0(0.6) & \nodata \\*
              & N & 3.9(2.2) & 1.3(0.5) & 0.2(0.4) & 0.7(0.4) & 4.8(0.8) & 0.3(0.7) & 2.4(0.7) & \nodata \\*
              & O & 4.6(3.3) & 0.3(0.6) & -0.1(0.6) & 0.8(0.6) & 4.7(1.1) & 1.0(0.8) & 2.4(0.8) & \nodata \\
\enddata

\tablenotetext{*}{The complete table will be available in the electronic version of the Astronomical Journal.}

\tablecomments{The adopted convention has positive EWs representing emission lines and negative EWs representing absorption, with errors placed in parentheses after the values. Where blends occur, the deblending has been performed in IRAF using the task {\it splot} as described in the text. The `Aperture' column denotes the extraction aperture, with `A' denoting the full 15\arcsec{} aperture, `N' denoting the 2\arcsec{} aperture, and `O' denoting the difference between the two. `M' denotes that the spectrum was obtained with the MX multifiber spectrometer, and no spatial information can be obtained. Such spectra are not flux calibrated. Values are not presented in two cases: 1) all or some portion of the defined continuum region lay off the CCD, and 2) blended lines that were too weak to be deblended.}

\end{deluxetable}

\begin{deluxetable}{c c c c c c c}
\tablecolumns{7}
\tablecaption{Balmer Line Fluxes for Cluster Radio Galaxies\tablenotemark{*}\label{tbl-LFs}}
\tabletypesize{\scriptsize}
\tablewidth{0pt}
\tablenum{5}
\tablehead{
\colhead{ID} & \colhead{Ap} & \colhead{H$\delta$} & \colhead{H$\gamma$} & 
\colhead{H$\beta$} & \colhead{H$\alpha$+[NII]} & \colhead{H$\alpha$} \cr
\colhead{} & \colhead{} & \colhead{$\lambda 4102$} & \colhead{$\lambda 4340$} & 
\colhead{$\lambda 4861$} & \colhead{$\lambda$6548,6563,6584} & \colhead{$\lambda 6563$}}
\startdata
015129+360356 & A &  0.69(0.05) &  2.85(0.15) &  5.40(0.18) & 29.3(0.69)  & 22.9(0.94) \\*
              & N &  0.00(0.04) &  0.23(0.02) &  0.43(0.02) &  5.07(0.14) &  3.52(0.17) \\
015254+360311 & A &  0.00(0.03) &  1.02(0.07) &  4.38(0.22) & 29.0(0.72)  & 23.1(0.99) \\*
              & N &  0.43(0.03) &  0.36(0.03) &  0.36(0.02) &  5.28(0.16) &  4.05(0.21) \\
015628+364807 & A &  4.74(0.12) &  5.40(0.12) &  7.03(0.09) & 46.2(0.40)  & 34.2(0.52) \\*
              & N &  1.22(0.03) &  1.55(0.03) &  2.67(0.04) & 19.1(0.16)  & 14.4(0.21) \\
021914+414258 & A &  2.72(0.10) &  6.93(0.12) & 14.8(0.26)  & 70.4(0.92)  & 52.2(1.19) \\*
              & N &  1.02(0.03) &  2.99(0.07) &  5.87(0.11) & 27.6(0.33)  & 18.4(0.38) \\
022004+411628 & A &  4.07(0.15) &  6.44(0.20) & 12.7(0.24)  & 53.0(0.84)  & 39.9(1.10) \\*
              & N &  1.57(0.05) &  2.16(0.07) &  4.86(0.11) & 19.7(0.33)  & 14.5(0.42) \\
\enddata

\tablenotetext{*}{The complete table will be available in the electronic version of the Astronomical Journal.}

\tablecomments{Only galaxies with emission lines are presented in this table. Values of 0.00 indicate that the line was not detected in emission. Where blends occur, the deblending has been performed in IRAF using the task {\it splot} as described in the text. The `Aperture' column denotes the extraction aperture, with `A' denoting the full 15\arcsec{} aperture and `N' denoting the 2\arcsec{} aperture (the off-nuclear flux is simply the difference between the two). Errors are based on the error in the continuum around the line, bootstrapped to the line region taking into account the relative sizes of the line and continuum regions. For deblended lines, a line extraction region of $20\mbox{\AA}$ was assumed.}

\end{deluxetable}

\begin{deluxetable}{c c c c c c c c c}
\tablecolumns{9}
\tablecaption{Forbidden Line Fluxes for Cluster Radio Galaxies\tablenotemark{*}\label{tbl-forLF}}
\tabletypesize{\tiny}
\tablewidth{0pt}
\tablenum{6}
\tablehead{
\colhead{ID} & \colhead{Ap} & \colhead{[OII]} & \colhead{[OIII]} & \colhead{[OIII]} & 
\colhead{[OI]} & \colhead{[NII]} & \colhead{[SII]} & \colhead{[SII]} \cr
\colhead{} & \colhead{} & \colhead{$\lambda 3727$} & \colhead{$\lambda 4959$} & 
\colhead{$\lambda 5007$} & \colhead{$\lambda 6300$} & \colhead{$\lambda 6584$} &
\colhead{$\lambda 6717$} & \colhead{$\lambda 6731$} \cr
 & & \multicolumn{7}{c}{all values $\times 10^{-15} erg~s^{-1}~cm^{-2}$}}
\startdata
015129+360356 & A &  1.27(0.36) &  1.51(0.05) &  0.62(0.02) &  0.95(0.03) &  8.60(0.35) &  3.32(0.26) &  3.48(0.28) \\*
              & N &  0.00(0.10) &  0.20(0.01) &  0.09(0.01) &  0.02(0.01) &  1.21(0.06) &  0.59(0.04) &  0.79(0.05) \\
015254+360311 & A &  4.81(1.40) &  0.34(0.02) &  0.48(0.02) &  0.26(0.01) &  7.02(0.30) &  1.58(0.08) &  0.00(0.00) \\*
              & N &  1.07(0.33) &  0.00(0.01) &  0.00(0.01) &  0.00(0.01) &  1.14(0.06) &  0.00(0.00) &  0.00(0.00) \\
015628+364807 & A &  6.73(0.54) &  2.04(0.03) &  0.00(0.00) &  2.53(0.03) & 16.4(0.25)  &  3.15(0.05) &  0.00(0.00) \\*
              & N &  2.21(0.17) &  1.54(0.02) &  0.24(0.01) &  0.90(0.01) &  6.38(0.10) &  3.09(0.05) &  0.00(0.00) \\
021914+414258 & A & 31.4(2.02)  &  3.11(0.05) &  7.61(0.12) &  1.25(0.03) & 16.5(0.38)  &  8.58(0.17) &  5.55(0.11) \\*
              & N & 10.8(0.74)  &  1.18(0.02) &  2.99(0.05) &  0.61(0.01) &  5.58(0.12) &  2.95(0.05) &  2.03(0.04) \\
022004+411628 & A & 29.0(2.38)  &  2.47(0.05) &  7.06(0.14) &  1.19(0.04) &  9.66(0.26) &  7.10(0.23) &  5.79(0.19) \\*
              & N &  8.66(0.69) &  1.13(0.03) &  2.75(0.07) &  0.42(0.01) &  3.46(0.10) &  2.44(0.07) &  1.91(0.06) \\
\enddata

\tablenotetext{*}{The complete table will be available in the electronic version of the Astronomical Journal.}

\tablecomments{Only galaxies with emission lines are presented in this table. Values of 0.00 indicate that the line was not detected in emission. Where blends occur, the deblending has been performed in IRAF using the task {\it splot} as described in the text. The `Aperture' column denotes the extraction aperture, with `A' denoting the full 15\arcsec{} aperture and `N' denoting the 2\arcsec{} aperture (the off-nuclear flux is simply the difference between the two). Errors are based on the error in the continuum around the line, bootstrapped to the line region taking into account the relative sizes of the line and continuum regions. For deblended lines, a line extraction region of $20\mbox{\AA}$ was assumed.}

\end{deluxetable}

\begin{deluxetable}{l c r l}
\tablecolumns{4}
\tablecaption{MORPHS Spectral Classification Scheme\label{tbl-morphs}}
\tablewidth{0pt}
\tablenum{7}
\tablehead{
\colhead{} & \colhead{EW [OII] 3727} & \colhead{EW H$\delta$} & \colhead{} \\
\colhead{Class} & \colhead{[$\mbox{\AA}$]} & \colhead{[$\mbox{\AA}$]} & \colhead{Comments}}
\startdata
k & Absent & $<3$ & Passive \\
k+a & Absent & 3--8 & Moderate Balmer absorption without emission \\
a+k & Absent & $\geq8$ & Strong Balmer absorption without emission \\
e(c) & Yes, $<40$ & $<4$ & Moderate Balmer absorption plus emission, spiral-like \\
e(a) & Yes & $\geq4$ & Strong Balmer absorption plus emission \\
e(b) & Yes, $\geq40$ & \nodata & Starburst \\
e(n) & \nodata & \nodata & AGN from broad lines or [OIII] 5007/H$\beta$ ratio
\enddata

\tablecomments{Adapted from \citet{dres1999}. The [OII] equivalent widths are for emission lines, the H$\delta$ are for absorption lines.}

\end{deluxetable}

\begin{deluxetable}{l l l l}
\tablecolumns{4}
\tablecaption{Galaxy Classifications\tablenotemark{*}\label{tbl-classes}}
\tablewidth{0pt}
\tablenum{8}
\tablehead{
\colhead{Cluster} & \colhead{Galaxy} & \colhead{Class} & \colhead{MORPHS}}
\startdata
A0262 & 015129+360356 & SF & k+a \\
A0262 & 015240+361016 & Ab+ & k \\
A0262 & 015246+360907 & Ab+ & k \\
A0262 & 015254+360311 & SF & k \\
A0262 & 015628+364807 & SF & e(c) \\
\enddata

\tablenotetext{*}{The complete table will be available in the electronic version of the Astronomical Journal.}

\tablenotetext{a}{The [OII] line was not available in this spectrum.}

\tablenotetext{b}{The wavelength coverage for the spectrum did not include the full blue continuum band for [OII], and hence no equivalent width or line flux is reported. However, the [OII] line is apparent in the spectrum.}

\tablecomments{Col. (1) Abell cluster to which galaxy belongs; col.(2) galaxy ID, from J2000 coordinates; col.(3) spectral class, where the codes are: SF -- star forming galaxy; SF? -- emission line galaxy presumed to be star forming, although line ratios are somewhat noisy; Abs -- absorption line spectrum with no apparent emission lines; Ab+ -- predominately absorption line spectrum, although with slight emission of [NII] and sometimes [SII]; Sey -- emission line galaxy shown to be a Seyfert in line ratio tests; Lin -- emission line galaxy shown to be a LINER in line ratio tests; Mix -- nuclear spectrum that of a Seyfert or LINER, with off nuclear spectrum showing star formation; col.(4) MORPHS classification (see Table \ref{tbl-morphs}), determined from full aperture spectrum.}

\end{deluxetable}

\begin{deluxetable}{l r c c c}
\tablecolumns{5}
\tablecaption{Observed Balmer Decrements\label{tbl-balmer}}
\tablewidth{0pt}
\tablenum{9}
\tablehead{
\colhead{Sample} & \colhead{N} & \colhead{$R_{all}$} & \colhead{$R_{nuc}$} & \colhead{$R_{off}$}}
\startdata
e(a) & 9 & 3.86$\pm$0.35 & 4.07$\pm$0.32 & 3.80$\pm$0.39 \\
e(b) & 7 & 4.15$\pm$0.44 & 4.32$\pm$0.50 & 4.28$\pm$0.56 \\
e(c) & 49 & 3.89$\pm$0.14 & 4.32$\pm$0.17 & 3.68$\pm$0.14 \\
e(a,b,c) & 65 & 3.92$\pm$0.13 & 4.29$\pm$0.15 & 3.76$\pm$0.13 \\
k, k+a & 23 & 4.39$\pm$0.32 & 6.16$\pm$0.40 & 3.96$\pm$0.35 \\
Net & 88 & 4.04$\pm$0.13 & 4.78$\pm$0.16 & 3.81$\pm0.14$ \\
\enddata

\tablecomments{Col. (1) Spectral classification in the MORPHS scheme; col.(2) Number of galaxies included. Note that these are only the galaxies for which long-slit spectra were obtained, since these have been properly flux-calibrated and adjusted for stellar Balmer absorption; col.(3) Balmer decrement for full 15\arcsec{} aperture; col.(4) Balmer decrement for nuclear 2\arcsec{} aperture; col.(5) Balmer decrement for off-nuclear aperture, defined as the difference between the 15\arcsec{} and 2\arcsec{} apertures.}
\end{deluxetable}

\begin{deluxetable}{l l c c c}
\tabletypesize{\small}
\tablecolumns{5}
\tablecaption{Core Radii\label{tbl-cores}}
\tablewidth{0pt}
\tablenum{10}
\tablehead{
\colhead{Class} & \colhead{L cut?} & \colhead{N} & \colhead{$r_c$} & \colhead{$\chi^2 / \nu$}}
\startdata
All & no & 445 & \phn600 & 3.38 \\
All & yes & 241 & \phn550 & 2.33 \\
SF  & no & 263 & \phn830 & 1.14 \\
SF  & yes & 125 & \phn950 & 0.59 \\
AGN & no & 182 & \phn180 & 2.55 \\
AGN & yes & 116 & \phn130 & 1.42 \\
OSP & no & \phn97 & \phn170 & 2.55 \\
OSP & yes & \phn59 & \phn170 & 1.00 \\
ELAGN & no & \phn21 & \phn900 & 0.58 \\
ELAGN & yes & \phn20 & 1130 & 0.63 \\
e(a) & no & \phn19 & \phn940 & 0.35 \\
e(a) & yes & \phn10 & 1160 & 0.30 \\
e(b) & no & \phn11 & \phn610 & 0.28 \\
e(b) & yes & \phn\phn7 & 1160 & 0.29 \\
e(c) & no & \phn85 & 1750 & 1.42 \\
e(c) & yes & \phn42 & 2120 & 1.50 \\
k & no & 100 & \phn220 & 1.31 \\
k & yes & \phn60 & \phn210 & 0.52 \\
k+a & no & \phn12 & \phn390 & 0.47 \\
k+a & yes & \phn\phn5 & \phn650 & 0.47 \\
SF:k,k+a & no & \phn27 & \phn450 & 0.65 \\
SF:k,k+a & yes & \phn14 & \phn470 & 0.21 \\
\enddata

\tablecomments{Col.(1) Classification of radio galaxies, where ``All'' signifies the full sample of radio galaxies, ``SF'' signifies those classified as star forming galaxies (including both those determined through spectroscopy and the FIR/radio correlation), and ``AGN'' signifies all AGN (including both those determined through spectroscopy and the FIR/radio correlation). The AGN are further broken down into those with absorption line spectra (``OSP'') and those with strong emission lines (``ELAGN''). The remaining classes are based exclusively on the optical spectroscopy and are defined in \citet{dres1999}; col.(2) designates whether any cutoff in radio luminosity was applied (see text); col.(3) number of galaxies contributing to fit; col.(4) best-fit core radius, in kpc; col.(5) reduced $\chi^2$ of the fit.}

\end{deluxetable}


\begin{thebibliography}{dummy}

\bibitem[Abell et al.(1989)]{abel1989} Abell, G.O., Corwin, H.G., \& Olowin, R.P. 1989, \apjs, 70, 1

\bibitem[Adami et al.(1998)]{adam1998} Adami, C., Mazure, A., Katgert, P., \& Biviano, A. 1998, \aap, 336, 63

\bibitem[Baldwin et al.(1981)]{bald1981} Baldwin, J., Phillips, M.M., \& Terlevich, R. 1981, \pasp, 93, 5

\bibitem[Balogh et al.(1998)]{balo1998} Balogh, M.L., Schade, D., Morris, S.L., Yee, H.K.C., Carlberg, R.G., \& Ellingson, E. 1998, \apj, 504, L75

\bibitem[Balogh, Navarro, \& Morris(2000)]{balo2000} Balogh, M.L., Navarro, J.F., \& Morris, S.L. 2000, \apj, 540, 113

\bibitem[Bekki(1999)]{bekk1999} Bekki, K. 1999, \apj, 510, L15

\bibitem[Bruzual(1983)]{bruz1983} Bruzual, A.G. 1983, \apj, 273, 105

\bibitem[Butcher \& Oemler(1978)]{butc1978} Butcher, H., \& Oemler, A. 1978, \apj, 219, 18

\bibitem[Butcher \& Oemler(1984)]{butc1984} Butcher, H., \& Oemler, A. 1984, \apj, 285, 426

\bibitem[Caldwell et al.(1993)]{cald1993} Caldwell, N., Rose, J.A., Sharples, R.M., Ellis, R.S., \& Bower, R.G. 1993, \aj, 106, 473

\bibitem[Caldwell \& Rose(1997)]{cald1997} Caldwell, N., \& Rose, J.A. 1997, \aj, 113, 492

\bibitem[Cardell, Clayton, \& Mathis(1989)]{card1989} Cardelli, J.A., Clayton, G.C., \& Mathis, J.S. 1989, \apj, 345, 245

\bibitem[Colless et al.(2001)]{coll2001} Colless, M. et al. 2001, in preparation (2dFGRS)

\bibitem[Condon(1992)]{cond1992} Condon, J.J. 1992, \araa, 30, 575

\bibitem[Condon et al.(1998)]{cond1998} Condon, J.J., Cotton, W.D., Greisen, E.W., Yin, Q.F., Perley, R.A., Taylor, G.B., \& Broderick, J.J. 1998, \aj, 115, 1693 (NVSS)

\bibitem[Couch \& Sharples(1987)]{couc1987} Couch, W.J., \& Sharples, R.M. 1987, \mnras, 229, 423

\bibitem[Coziol et al.(1998)]{cozi1998} Coziol, R., Ribeiro, A.L., de Carvalho, R.R., \& Capelato, H.V. 1998, \apj, 493, 563

\bibitem[Dressler \& Gunn(1983)]{dres1983} Dressler, A., \& Gunn, J.E. 1983, \apj, 270, 7

\bibitem[Dressler \& Shectman(1987)]{drsh1987} Dressler, A., \& Shectman, S.A. 1987, \aj, 94, 899

\bibitem[Dressler \& Gunn(1992)]{dres1992} Dressler, A., \& Gunn, J.E. 1992, \apjs, 78, 1

\bibitem[Dressler et al.(1997)]{dres1997} Dressler, A., Oemler, A. Jr., Couch, W.J., Smail, I., Ellis, R.S., Barger, A., Butcher, H., Poggianti, B.M., \& Sharples, R.M. (1997), \apj, 490, 577

\bibitem[Dressler et al.(1999)]{dres1999} Dressler, A., Smail, I.R., Poggianti, B.M., Butcher, H., Couch, W.J., Ellis, R.S., \& Oemler, A. Jr. 1999, \apjs, 122, 51

\bibitem[Dwarakanath \& Owen(1999)]{dwar1999} Dwarakanath, K.S., \& Owen, F.N. 1999, \aj, 118, 625

\bibitem[Ellingson et al.(2001)]{elli2001} Ellingson, E., Lin, H., Yee, H. K. C., \& Carlberg, R. G. 2001, \apj, 547, 609

\bibitem[Fujita \& Nagashima(1999)]{fuji1999} Fujita, Y., \& Nagashima, M. 1999, \apj, 516, 619

\bibitem[Gonzalez Delgado et al.(1999)]{gonz1999} Gonzalez Delgado, R.M., Leitherer, C., \& Heckman, T. 1999, \apjs, 125, 489

\bibitem[Gunn \& Gott(1972)]{gunn1972} Gunn, J.E., \& Gott, J.R. 1972, \apj, 176, 1

\bibitem[Hashimoto et al.(1998)]{hash1998} Hashimoto, Y., Oemler, A., Lin, H., and Tucker, D.L. 1998, \apj, 499, 589

\bibitem[Heckman(1980)]{heck1980} Heckman, T. 1980, \aap, 87, 152

\bibitem[Helou, Soifer, \& Rowan-Robinson(1985)]{helo1985} Helou, G., Soifer, B.T., \& Rowan-Robinson, M. 1985, \apj, 298, L11

\bibitem[Henriksen \& Byrd(1996)]{henr1996} Henriksen, M.J., \& Byrd, G. 1996, \apj, 459, 82

\bibitem[Hickson(1982)]{hick1982} Hickson, P. 1982, \apj, 255, 382

\bibitem[Hill \& Lesser(1986)]{hill1986} Hill, J.M., \& Lesser, M.P. 1986, Proceedings of SPIE Conference on Instrumentation in Astronomy VI, SPIE 627, 303

\bibitem[Hill \& Lesser(1988)]{hill1988} Hill, J.M., \& Lesser, M.P. 1988, Instrumentation for Ground-Based Astronomy: Present \& Future, edited by Lloyd B. Robinson (Springer:New York), p. 233

\bibitem[Ho(1996)]{ho1996} Ho, L.C. 1996, in {\it The Physics of LINERs in View of Recent Observations} ASP???

\bibitem[Ho(1999)]{ho1999} Ho, L.C. 1999, \apj, 510, 631

\bibitem[Keel(1983)]{keel1983} Keel, W.C. 1983, \apj, 269, 466

\bibitem[Kennicutt(1992a)]{kenn1992} Kennicutt, R.C. 1992, \apjs, 79, 255

\bibitem[Kennicutt(1992b)]{kenn92b} Kennicutt, R.C. 1992, \apj, 388, 310

\bibitem[King(1966)]{king1966} King, I.R. 1966, \aj, 71, 64

\bibitem[Liu \& Kennicutt(1995)]{liu1995} Liu, C.T., \& Kennicutt, R.C. 1995, \apjs, 100, 325

\bibitem[Lubin(1996)]{lubi1996} Lubin, L.M. 1996, \aj, 112, 23

\bibitem[Machalski \& Condon(1999)]{mach1999} Machalski, J., \& Condon, J.J. 1999, \apjs, 123, 41 (MC99)

\bibitem[Machalski \& Godlowski(2000)]{mach2000} Machalski, J., \& Godlowski, W. 2000, \aap, 360, 463 (MG00)

\bibitem[Margoniner \& de Carvalho(2000)]{marg2000} Margoniner, V.E., \& de Carvalho, R.R. 2000, \aj, 119, 1562

\bibitem[Margoniner et al.(2001)]{marg2001} Margoniner, V.E., de Carvalho, R.R., Gal, R.R., \& Djorgovski, S.G. 2001, \apj, 548, L143

\bibitem[Miller \& Owen(2001a)]{mill2001} Miller, N.A., \& Owen, F.N. 2001, \apjs, 134, 355 (Paper I)

\bibitem[Miller \& Owen(2001b)]{mil2001b} Miller, N.A., \& Owen, F.N. 2001, \aj, 121, 1903 (Paper II)

\bibitem[Miller \& Owen(2001c)]{mil2001c} Miller, N.A., \& Owen, F.N. 2001, \apj, 554, L25

\bibitem[Moore, Lake, \& Katz(1998)]{moor1998} Moore, B., Lake, G., \& Katz, N. 1998, \apj, 495, 139

\bibitem[Morrison \& Owen(2002)]{morr1999} Morrison, G.E., \& Owen, F.N. 2002, \aj, in submission

\bibitem[Moss \& Whittle(2000)]{moss2000} Moss, C., \& Whittle, M. 2000, \mnras, 317, 667

\bibitem[Osterbrock(1989)]{oste1989} Osterbrock, D. 1989, {\it Astrophysics of Gaseous Nebulae and Active Galactic Nuclei} (Mill Valley, CA:University Science Books)

\bibitem[Owen \& Laing(1989)]{owen1989} Owen, F.N., \& Laing, R.A. 1989, \mnras, 238, 357

\bibitem[Owen et al.(1999)]{owen1999} Owen, F.N., Ledlow, M.J., Keel, W.C., \& Morrison, G.E. 1999, \aj, 118, 633

\bibitem[Phillips et al.(1986)]{phil1986} Phillips, M.M., Jenkins, C.R., Dopita, M.A., Sadler, E.M., \& Binette, L. 1986, \aj, 91, 1062

\bibitem[Poggianti \& Barbaro(1997)]{pogg1997} Poggianti, B.M., \& Barbaro, G. 1997, \aap, 325, 1025

\bibitem[Poggianti et al.(1999)]{pogg1999} Poggianti, B.M., Smail, I.R., Dressler, A., Couch, W.J., Barger, A.J., Butcher, H., Ellis, R.S., \& Oemler, A. Jr. 1999, \apj, 518, 576

\bibitem[Poggianti \& Wu(2000)]{pogg2000} Poggianti, B.M., \& Wu, H. 2000, \apj, 529, 157

\bibitem[Quilis, Moore, \& Bower(2000)]{quil2000} Quilis, V., Moore, B., \& Bower, R. 2000, Science, 288, 1617

\bibitem[Rakos \& Schombert(1995)]{rako1995} Rakos, K.D., \& Schombert, J.M. 1995, \apj, 439, 47

\bibitem[Rakos, Odell, \& Schombert(1997)]{rako1997} Rakos, K.D., Odell, A.P., \& Schombert, J.M. 1997, \apj, 490, 194

\bibitem[Rose et al.(2001)]{rose2001} Rose, J.A., Gaba, A.E., Caldwell, N., \& Chaboyer, B. 2001, \aj, 121, 793

\bibitem[Sadler et al.(1999)]{sadl1999} Sadler, E.M., McIntyre, V.J., Jackson, C.A., \& Cannon, R.D. 1999, PASA, 16, 247

\bibitem[Sadler et al.(2002)]{sadl2002} Sadler, E.M., Jackson, C.A., Cannon, R.D., McIntyre, V.J., Murphy, T., Baugh, C.M., Bland-Hawthorn, J., Bridges, T., Cole, S., Colless, M., Collins, C., Couch, W., Dalton, G., De Propris, R., Driver, S.P., Efstathiou, G., Ellis, R.S., Frenk, C.S., Glazebrook, K., Lahav, O., Lewis, I., Lumsden, S., Maddox, S., Madgwick, D., Norberg, P., Peacock, J.A., Peterson, B.A., Sutherland, W., \& Taylor, K. 2001, \mnras, 329, 227

\bibitem[Shectman et al.(1996)]{shec1996} Shectman, S.A., Landy, S.D., Oemler, A., Tucker, D.L., Lin, H., Kirshner, R.P., \& Schechter, P.L. 1996, \apj, 470, 172 (LCRS)

\bibitem[Shioya, Bekki, \& Couch(2001)]{shio2001} Shioya, Y., Bekki, K., \& Couch, W.J. 2001, \apj, 558, 42

\bibitem[Shioya et al.(2002)]{shio2002} Shioya, Y., Bekki, K., Couch, W.J., \& De Propis, R. 2002, \apj, 565, 233

\bibitem[Smail et al.(1999)]{smai1999} Smail, I., Morrison, G.E., Gray, M.E., Owen, F.N., Ivison, R.J., Kneib, J.-P., \& Ellis, R.S. 1999, \apj, 525, 609

\bibitem[Veilleux \& Osterbrock(1987)]{veil1987} Veilleux, S., \& Osterbrock, D.E. 1987, \apjs, 63, 295

\bibitem[Yun, Reddy, \& Condon(2001)]{yun2001} Yun, M.S., Reddy, N.A., \& Condon, J.J. 2001, \apj, 554, 803

\bibitem[Zabludoff et al.(1996)]{zabl1996} Zabludoff, A.I., Zaritsky, D., Lin, H., Tucker, D., Hashimoto, Y., Shectman, S.A., Oemler, A. Jr., \& Kirshner, R.P. 1996, \apj, 466, 104

\end{thebibliography}
\end{document}